\documentclass[12pt]{article}
\usepackage{latexsym}
\usepackage{amsmath,amsfonts}
\usepackage{times}
\allowdisplaybreaks[4]

\catcode`\@=11

\newcount\hour
\newcount\minute
\newtoks\amorpm \hour=\time\divide\hour by 60\minute
=\time{\multiply\hour by 60 \global\advance\minute by-\hour}
\edef\standardtime{{\ifnum\hour<12 \global\amorpm={am}%
        \else\global\amorpm={pm}\advance\hour by-12 \fi
        \ifnum\hour=0 \hour=12 \fi
        \number\hour:\ifnum\minute<10
        0\fi\number\minute\the\amorpm}}
\edef\militarytime{\number\hour:\ifnum\minute<10
0\fi\number\minute}

\def\draftlabel#1{{\@bsphack\if@filesw {\let\thepage\relax
   \xdef\@gtempa{\write\@auxout{\string
      \newlabel{#1}{{\@currentlabel}{\thepage}}}}}\@gtempa
   \if@nobreak \ifvmode\nobreak\fi\fi\fi\@esphack}
        \gdef\@eqnlabel{#1}}
\def\@eqnlabel{}
\def\@vacuum{}
\def\marginnote#1{}
\def\draftmarginnote#1{\marginpar{\raggedright\scriptsize\tt#1}}
\def\draft{
        \pagestyle{plain}
        \overfullrule=2pt
        \oddsidemargin -.5truein
        \def\@oddhead{\sl \phantom{\today\quad\militarytime} \hfil
        \smash{\Large\sl DRAFT} \hfil \today\quad\militarytime}
        \let\@evenhead\@oddhead
        \let\label=\draftlabel
        \let\marginnote=\draftmarginnote
        \def\ps@empty{\let\@mkboth\@gobbletwo
        \def\@oddfoot{\hfil \smash{\Large\sl DRAFT} \hfil}
        \let\@evenfoot\@oddhead}
        \def\@eqnnum{(\theequation)\rlap{\kern\marginparsep\tt\@eqnlabel}%
        \global\let\@eqnlabel\@vacuum}  }

\newcommand{\rf}[1]{(\ref{#1})}
\renewcommand{\theequation}{\thesection.\arabic{equation}}
\renewcommand{\thefootnote}{\fnsymbol{footnote}}
\newcommand{\newsection}{    
\setcounter{equation}{0}\section}

\def\appendix#1{\addtocounter{section}{1}\setcounter{equation}{0}
\renewcommand{\thesection}{\Alph{section}}
\section*{Appendix \thesection\protect\indent \parbox[t]{11.15cm}{#1}}
\addcontentsline{toc}{section}{Appendix \thesection\ \ \ #1}}

\def\asf{{\sf a}}
\def\bsf{{\sf b}}
\def\csf{{\sf c}}
\def\esf{{\sf e}}
\def\Nsf{{\sf N}}

\def\nline{\,\nabla\kern -0.7em\raise0.2ex\hbox{/}\,\,}
\def\yline{\,y\kern -0.47em /}
\def\aline{\,a\kern -0.49em /}
\def\parline{\,\partial\kern -0.55em /\,\,}

\newcommand{\Eo}{\mathbb{E}}

\newcommand{\Mo}{\mathbb{M}}
\newcommand{\No}{\mathbb{N}}
\newcommand{\Po}{\mathbb{P}}

\newcommand{\Zo}{\mathbb{Z}}

\def\be{\begin{equation}}
\def\ee{\end{equation}}
\def\beq{\begin{eqnarray}}
\def\eeq{\end{eqnarray}}

\def\Rsm{{\scriptscriptstyle R}}
\def\Lsm{{\scriptscriptstyle L}}

\def\smpt{{\scriptscriptstyle [2]}}
\def\smp3{{\scriptscriptstyle [3]}}

\def\smpn{{\scriptscriptstyle [n]}}

\def\sbf{{\bf s}}
\def\Jbf{{\bf J}}
\def\Ebf{{\bf E}}

\def\Mbf{{\bf M}}
\def\Pbf{{\bf P}}

\def\Xbf{{\bf X}}
\def\Vbf{{\bf V}}

\def\ibf{{\bf i}}
\def\iibf{{\bf ii}}
\def\iiibf{{\bf iii}}
\def\ivbf{{\bf iv}}

\def\LL{{\cal L}}

\def\PP{{\cal P}}

\def\Nsf{{\sf N}}

\def\half{\frac{1}{2}}

\def\Cb{{\bar{C}}}

\def\Vb{{\bar{V}}}
\def\vb{{\bar{v}}}

\def\irm{{\rm i}}

\def\dyn{{\rm dyn}}
\def\minrm{{\rm min}}

\def\diff{{\rm diff}}
\def\GP{{\rm GP}}
\def\cov{{\rm cov}}
\def\lc{{\rm lc}}

\def\betach{\check{\beta}}

\begin{document}


\begin{flushright}
FIAN-TD-2019-15  \ \ \ \ \ \\
arXiv: 1905.11357V3
\end{flushright}

\vspace{1cm}

\begin{center}

{\Large \bf Cubic interaction vertices for N=1 arbitrary spin massless

\medskip
supermultiplets in flat space}

\vspace{2.5cm}

R.R. Metsaev\footnote{ E-mail: metsaev@lpi.ru }

\vspace{1cm}

{\it Department of Theoretical Physics, P.N. Lebedev Physical
Institute, \\ Leninsky prospect 53,  Moscow 119991, Russia }

\vspace{3cm}

{\bf Abstract}

\end{center}

In the framework of light-cone gauge formulation, massless arbitrary spin  N=1 supermultiplets in four-dimensional flat space are considered. We study both the integer spin and half-integer spin supermultiplets.  For such supermultiplets, formulation in terms of unconstrained light-cone gauge superfields defined in momentum superspace is used. Superfield representation for all cubic interaction vertices of the supermultiplets is obtained. Representation of the cubic vertices in terms of component fields is derived. Realization of relativistic symmetries of N=1 Poincar\'e superalgebra on space of interacting superfields is also found.

\vspace{3cm}


\newpage
\renewcommand{\thefootnote}{\arabic{footnote}}
\setcounter{footnote}{0}

\section{ \large Introduction}

The light-cone gauge approach \cite{Dirac:1949cp}
offers considerable simplifications of approaches  to
many problems of quantum field theory and superstring. This approach
hides Lorentz symmetries but eventually turns out to be effective.
Exploring this approach for the analysis of ultraviolet finiteness of $N=4$ Yang-Mills theory may be found in Refs.\cite{Brink:1982wv,Mandelstam:1982cb}.
Light-cone gauge superstring field theories are studied in Refs.\cite{Green:1983hw}, while string bit models for superstring and super $p$-branes in the framework of light-cone gauge formulation are considered in Ref.\cite{Bergman:1995wh}  and \cite{deWit:1988wri} respectively. Application of light-cone gauge formalism for studying the interacting continuous-spin fields in flat space may be found in Refs.\cite{Metsaev:2017cuz,Metsaev:2018moa}.
Various applications of light-cone gauge approach to field theory like QCD are discussed in \cite{Brodsky:2013dca}.
Methods for building Lorentz covariant formulation by using light-cone gauge formulation  are investigated in Ref.\cite{Siegel:1988yz}.
In the framework of light-cone gauge approach, study of free continuous-spin field in AdS space may be found in Refs.\cite{Metsaev:2017myp,Metsaev:2019opn}.

One interesting application of light-cone gauge approach is a higher-spin massless field theory.
In Refs.\cite{Bengtsson:1983pd,Bengtsson:1983pg}, a wide class of cubic interaction vertices for higher-spin massless fields in $4d$ flat space was constructed, while the full list of cubic interaction vertices for arbitrary spin massless fields in $4d$ flat space was obtained in Ref.\cite{Bengtsson:1986kh}.%
\footnote{ Generalization of results in Ref.\cite{Bengtsson:1986kh} to the case of massless and massive arbitrary spin bosonic and fermionic fields in $R^{d-1,1}$, $d$-arbitrary, may be found in Refs.\cite{Metsaev:2005ar,Metsaev:2007rn}
}
Our aim in this paper is to provide the full list of cubic interaction vertices for $N=1$ arbitrary integer and half-integer spin supermultiplets in the flat $4d$ space. Doing so, we provide, among other things, the supersymmetric extension for all cubic interaction vertices for massless bosonic fields in the $4d$ flat space presented in Ref.\cite{Bengtsson:1986kh}.%
\footnote{ In the framework of light-cone superspace formalism, a scalar superfield that describes arbitrary $N$-extended supermultiplets and involves fields with helicities  $ - \frac{1}{4} N \leq \lambda \leq \frac{1}{4} N$ ($\frac{1}{4}N$ -integer) was studied in Ref.\cite{Bengtsson:1983pg}. For such scalar superfield, a cubic vertex that involves $\frac{1}{4} N$ derivatives was obtained in Ref.\cite{Bengtsson:1983pg}.}
To this end we use superfields defined in light-cone momentum superspace. The light-cone momentum superspace has successfully been used in many interesting studies of superstring and supergravity  theories. For example, we mention the use of the momentum superspace in superstring field theories in Refs.\cite{Green:1983hw} and 10d extended supergravity in Ref.\cite{Green:1982tk}.%
\footnote{ Recent interesting discussion of 10d Yang-Mills theory in light-cone superspace may be found in Ref.\cite{Ananth:2015tsa}.}
The momentum superspace was also adapted for the studying light-cone gauge 11d supergravity in Ref.\cite{Metsaev:2004wv}. Using the momentum superspace, we collect the $N=1$ integer and half-integer spin massless supermultiplets into a suitable unconstrained superfields and use such superfields for building cubic interaction vertices. It is the use of the light-cone gauge unconstrained superfields that allows us to build a simple representation for all cubic interaction vertices of the $N=1$ integer and half-integer spin massless supermultiplets and to provide the full classification of such cubic vertices.

This paper is organized as follows.

In Sec.\ref{sec-02}, we review the well known description of $N=1$ integer spin and half-integer spin supermultiplets in terms of light-cone gauge components fields. We introduce the field content and describe a realization of the Poincar\'e superalgebra on space of the component fields.

In Sec.\ref{sec-superfield},  we introduce a momentum superspace and describe light-cone gauge unconstrained superfields defined in such superspace.
Also we describe a realization of the Poincar\'e superalgebra on space of our light-cone gauge unconstrained superfields.

In Sec.\ref{sec-npoint}, we describe a general structure of $n$-point interaction vertices. Namely, we present restrictions imposed by kinematical symmetries of the Poincar\'e superalgebra on $n$-point interaction vertices.

In Sec.\ref{sec-05}, we study cubic vertices. First, we present restrictions imposed by kinematical and dynamical symmetries of the Poincar\'e superalgebra on cubic vertices. Second, we formulate light-cone gauge dynamical principle and present complete system of equations required to determine the cubic vertices uniquely.

In Sec.\ref{sec-06}, we present our general solution for all cubic vertices which describe interactions of arbitrary spin massless supermultiplets. First, we present superfield form of the cubic vertices. After that, we discuss the cubic vertices in terms of the component fields and provide the full classification of the cubic vertices for integer and half-integer arbitrary spin supermultiplets.

Sec.\ref{concl} is devoted to our conclusions.

In Appendix A, we describe our basic notation and conventions for Grassmann algebra we use in this paper. In Appendix B, we discuss properties of our unconstrained superfields.
In appendix C, we present some details of the derivation of the cubic vertices.

\newsection{ \large Light-cone gauge formulation of free massless $N=1$ supermultiplets }\label{sec-02}

\noindent {\bf Poincar\'e superalgebra in light-cone frame}. A method suggested in Ref.\cite{Dirac:1949cp} tells us that the problem of finding a light-cone gauge dynamical system  amounts to a problem of finding a light cone gauge solution for commutation relations of a symmetry algebra.
For supersymmetric theories in the flat space $R^{3,1}$, basic symmetries are associated with the Poincar\'e superalgebra. Therefore, in this section, we review a realization of the Poincar\'e superalgebra on a space of massless supermultiplets
and present well known formulation of free $N=1$ supersymmetric  multiplets in terms of the light-cone gauge component fields.

For the flat space $R^{3,1}$, the Poincar\'e superalgebra is spanned by the four translation generators $P^\mu$, the six generators of the $so(3,1)$ Lorentz algebra $J^{\mu\nu}$, and four
Majorana supercharges $Q^\alpha$. We assume the following (anti)commutators:
\beq
\label{11052019-man-01} && [P^\mu,\,J^{\nu\rho}]=\eta^{\mu\nu} P^\rho - \eta^{\mu\rho} P^\nu\,,
\qquad {} [J^{\mu\nu},\,J^{\rho\sigma}] = \eta^{\nu\rho} J^{\mu\sigma} + 3\hbox{ terms}\,,
\\
\label{11052019-man-02} && [Q,J^{\mu\nu}] = \half \gamma^{\mu\nu} Q\,, \hspace{2.5cm} \{ Q^\alpha, Q^\beta\} = - (\gamma^\mu C^{-1})^{\alpha\beta} P^\mu\,,
\eeq
where $\eta^{\mu\nu}$ ia the mostly positive Minkowski metric. We do not present an explicit form of $\gamma^\mu$-matrices and charge conjugation $C$-matrix because throughout this paper we use only light-cone form of (anti)commutators \rf{11052019-man-01},\rf{11052019-man-02}.

In place of the Lorentz basis coordinates $x^\mu$, $\mu=0,1,2,3$, we introduce the light-cone basis coordinates $x^\pm$, $x^\Rsm$, $x^\Lsm$ defined as
\be \label{11052019-man-03}
x^\pm \equiv \frac{1}{\sqrt{2}}(x^3  \pm x^0)\,,\qquad
\qquad x^\Rsm \equiv \frac{1}{\sqrt{2}}(x^1 + \irm x^2)\,,\qquad x^\Lsm \equiv \frac{1}{\sqrt{2}}(x^1 - \irm x^2)\,,
\ee
where the coordinate $x^+$ is considered as an evolution parameter. In the light-cone basis \rf{11052019-man-03}, the $so(3,1)$ Lorentz algebra vector $X^\mu$ is decomposed as $X^+,X^-,X^\Rsm$, $X^\Lsm$, while a scalar product of the $so(3,1)$ Lorentz algebra vectors $X^\mu$ and $Y^\mu$ is decomposed as
\be  \label{11052019-man-04}
\eta_{\mu\nu}X^\mu Y^\nu = X^+Y^- + X^-Y^+ + X^\Rsm Y^\Lsm + X^\Lsm Y^\Rsm\,.
\ee
From \rf{11052019-man-04}, we learn that, in the light-cone basis, non-vanishing elements of the flat metric are given by $\eta_{+-} = \eta_{-+}=1$, $\eta_{\Rsm\Lsm} = \eta_{\Lsm\Rsm} = 1$. This implies that the covariant and contravariant components of vector $X^\mu$ are related as $X^+=X_-$, $X^-=X_+$, $X^\Rsm=X_\Lsm$, $X^\Lsm=X_\Rsm$.

In light-cone basis \rf{11052019-man-03}, generators of the Poincar\'e superalgebra are separated into two groups:

{\small
\beq
\label{11052019-man-05} && \hspace{-1.6cm}
P^+,\quad
P^\Rsm,\hspace{0.6cm}
P^\Lsm,\hspace{0.5cm}
J^{+\Rsm},\quad
J^{+\Lsm},\quad
J^{+-},\quad
J^{\Rsm\Lsm}, \quad
Q^{+\Rsm}, \quad
Q^{+\Lsm}, \quad
\hbox{ kinematical generators};
\\
\label{11052019-man-06}  && \hspace{-1.6cm}
P^-, \quad
J^{-\Rsm}, \quad
J^{-\Lsm}, \quad
Q^{-\Rsm}, \quad
Q^{-\Lsm}, \quad \hspace{5cm}
\hbox{ dynamical generators}.
\eeq
}
We recall that, for $x^+=0$, in a field realization, generators \rf{11052019-man-05} are quadratic in fields%
\footnote{ With exception of $J^{+-}$, generators \rf{11052019-man-05} are also quadratic in fields when $x^+ \ne 0 $.  The $J^{+-}$ takes the form $J^{+-} = G_0 + \irm x^+ P^-$, where  $G_0$ is quadratic in fields, while $P^-$ involves quadratic and higher order terms in fields.},
while, generators \rf{11052019-man-06} involve quadratic and higher order terms in fields.

In the light-cone basis, commutators of the Poincar\'e algebra are obtained from \rf{11052019-man-01} simply by using the flat metric $\eta^{\mu\nu}$ which has non-vanishing elements $\eta^{+-}=\eta^{-+}=1$, $\eta^{\Rsm\Lsm}=\eta^{\Lsm\Rsm}=1$. We now present the light-cone form of the
(anti)commutators given in \rf{11052019-man-02},
\beq
&& [J^{+-},Q^{\pm \Rsm}] = \pm \half Q^{\pm \Rsm}\,,\qquad [J^{+-},Q^{\pm \Lsm}] = \pm \half Q^{\pm \Lsm}\,,
\\
&& [J^{\Rsm\Lsm},Q^{\pm \Rsm}] = \half Q^{\pm \Rsm}\,,\qquad \quad [J^{\Rsm\Lsm},Q^{\pm \Lsm}] = - \half Q^{\pm \Lsm}\,,
\\
&& [Q^{-\Rsm},J^{+\Lsm}] = - Q^{+\Lsm}\,, \hspace{1.3cm} [Q^{-\Lsm},J^{+\Rsm}] = - Q^{+\Rsm}\,,
\\
&& [Q^{+\Rsm},J^{-\Lsm}] =   Q^{-\Lsm}\,, \hspace{1.7cm} [Q^{+\Lsm},J^{-\Rsm}] =  Q^{-\Rsm}\,,
\\[10pt]
&& \{ Q^{+\Rsm},Q^{+\Lsm} \} = P^+\,, \hspace{1.6cm} \{ Q^{-\Rsm},Q^{-\Lsm} \} = - P^-\,,
\\
&&  \{ Q^{+\Rsm},Q^{-\Rsm} \} = P^\Rsm\,, \hspace{1.6cm}   \{Q^{+\Lsm},Q^{-\Lsm}\} = P^\Lsm\,.
\eeq

Hermitian conjugation rules for the generators are assumed to be as follows
\beq
&& \hspace{-1.4cm}
P^{\pm \dagger} = P^\pm, \qquad \ \
P^{\Rsm\dagger} = P^\Lsm, \qquad
J^{\Rsm\Lsm\dagger} =  J^{\Rsm\Lsm}\,,\quad
J^{+-\dagger} = - J^{+-}, \quad
J^{\pm \Rsm\dagger} = -J^{\pm \Lsm}\,,
\nonumber\\
&& \hspace{-1.4cm}
Q^{+\Rsm\dagger} = Q^{+\Lsm}\,, \hspace{0.6cm}  Q^{-\Rsm\dagger} = Q^{-\Lsm}\,. \qquad
\eeq
In order to provide a field theoretical realization of generators of the Poincar\'e superalgebra on massless fields, we use a light-cone gauge formulation. To this end we start with a description of field content we use in this paper and review the well known light-cone gauge formulation of arbitrary spin massless fields.

\noindent {\bf Field content}. To discuss supersymmetric field theories we use light-cone gauge massless fields considered in helicity basis. First, using a label $\lambda$ to denote a helicity of a massless field, we introduce the following set of complex-valued massless fields:
\beq
\label{15052019-man02-01} &&\hspace{-2.5cm}  \hbox{bosonic fields:} \hspace{0.6cm} \phi_\lambda\,, \hspace{0.3cm}
\lambda =\pm 1\,,\pm 2,\ldots, \pm \infty,  \hspace{0.8cm} \psi_\lambda\,, \quad \lambda =\pm 1\,,\pm 2,\ldots, \pm \infty;
\\
\label{15052019-man02-02} && \hspace{-2.5cm} \hbox{fermionic fields:} \hspace{0.3cm} \phi_\lambda\,, \hspace{0.2cm} \lambda  = \hbox{$ \pm\, \frac{1}{2},\,\pm\,  \frac{3}{2}, \ldots, \pm \infty $},  \hspace{0.7cm} \psi_\lambda\,, \hspace{0.3cm} \lambda  = \hbox{$ \pm\, \frac{3}{2}, \pm \frac{5}{2}, \ldots, \pm \infty $}, \qquad
\eeq
where fields \rf{15052019-man02-01},\rf{15052019-man02-02} depend on space time-coordinates $x^\pm$, $x^{\Rsm,\Lsm}$ \rf{11052019-man-03}. Fields \rf{15052019-man02-01},\rf{15052019-man02-02} satisfy the following hermitian conjugation rules
\be
\label{15052019-man02-03} \phi_\lambda^\dagger(x) = \phi_{-\lambda}(x)\,,\qquad \psi_\lambda^\dagger(x) = \psi_{-\lambda}(x)\,.
\ee
We collect fields  \rf{15052019-man02-01},\rf{15052019-man02-02} into integer and half-integer spin supermultiplets given by
\beq
\label{15052019-man02-04} && \hspace{-1.2cm} (\phi_s\,, \phi_{s-\half})\qquad (\phi_{-s}\,, \phi_{-s+\half})\qquad \hbox{spin-$s$ supermultiplets},\hspace{2cm} s =1,2,\ldots,\infty;\quad
\\
\label{15052019-man02-05} && \hspace{-1.2cm} (\psi_{s+\half}\,, \psi_s)\qquad (\psi_{-s-\half}\,, \psi_{-s})\qquad \hbox{spin-$(s+\half)$ supermultiplets},\hspace{0.9cm} s =1,2,\ldots,\infty;\quad
\eeq
Second, by analogy with  \rf{15052019-man02-05}, we introduce the spin-$\half$ massless supermultiplets
\beq
\label{15052019-man02-06}  && (\psi_\half\,, \psi_0)\qquad (\psi_{-\half}\,, \psi_{-0})\qquad \hbox{spin-$\half$ supermultiplets},
\\
\label{15052019-man02-07}
&& \psi_0^\dagger(x) = \psi_{-0}(x)\,, \qquad \psi_\half^\dagger(x) = \psi_{-\half}(x)\,,
\eeq
where, in \rf{15052019-man02-07}, we fix the hermitian conjugation rules for two complex-valued scalar fields $\psi_0$, $\psi_{-0}$ and two complex-valued spin-$\half$ fermionic fields $\psi_{1/2}$, $\psi_{-1/2}$.

From \rf{15052019-man02-03}-\rf{15052019-man02-07}, we see that the supermultiplet $\phi_s,\phi_{s-\half}$ is hermitian conjugated to the supermultiplet  $\phi_{-s},\phi_{-s+\half}$, while the supermultiplet  $\psi_{s+\half},\psi_s$ is hermitian conjugated to the supermultiplet $\psi_{-s-\half},\psi_{-s}$. For supermultiplets $\phi_s,\phi_{s-\half}$ and $\phi_{-s},\phi_{-s+\half}$ \rf{15052019-man02-04}, we use the shortcut {\small $(s,s-\half)$}, $s=1,2,\ldots,\infty$, while, for supermultiplets $\psi_{s+\half}, \psi_s$ and  $\psi_{-s-\half},\psi_{-s}$ \rf{15052019-man02-05},\rf{15052019-man02-06}, we use the shortcut {\small$(s+\half,s)$}, $s=0,1,\ldots,\infty$.

{\it Fields in \rf{15052019-man02-04}-\rf{15052019-man02-06} constitute the  field content in our approach}. In our field content, each helicity occurs twice. Our motivation for the use of such field content is discussed in Sec.6.

In what follows, we prefer to use fields which obtained from the ones in \rf{15052019-man02-01}-\rf{15052019-man02-06} by using the Fourier transform with respect to the coordinates $x^-$, $x^\Rsm$, and $x^\Lsm$,
\beq
\label{15052019-man02-09} && \phi_\lambda(x) = \int \frac{ d^3p }{ (2\pi)^{3/2} } e^{\irm(\beta x^- + p^\Rsm x^\Lsm  + p^\Lsm x^\Rsm)} \phi_\lambda(x^+,p)\,,
\nonumber\\
\label{15052019-man02-10} && \psi_\lambda(x) = \int \frac{ d^3p }{ (2\pi)^{3/2} } e^{\irm (\beta x^- + p^\Rsm x^\Lsm  + p^\Lsm x^\Rsm)} \psi_\lambda(x^+,p)\,, \qquad d^3p \equiv d\beta dp^\Rsm dp^\Lsm\,,
\eeq
where the argument $p$ of fields $\phi_\lambda(x^+,p)$, $\psi_\lambda(x^+,p)$ stands for the momenta $\beta$, $p^\Rsm$, $p^\Lsm$. In terms of the fields $\phi_\lambda(x^+,p)$, $\psi_\lambda(x^+,p)$ , the hermicity conditions \rf{15052019-man02-03},\rf{15052019-man02-07} take the form
\be \label{15052019-man02-11}
\phi_\lambda^\dagger(p) = \phi_{-\lambda}(-p)\,,\qquad \psi_\lambda^\dagger(p) = \psi_{-\lambda}(-p)\,,
\ee
where in \rf{15052019-man02-11} and below dependence of the fields on the light-cone time $x^+$ is implicit.

\noindent {\bf Field-theoretical realization of the Poincar\'e superalgebra}. We now review a field theoretical realization of the Poincar\'e superalgebra on the space of massless supermultiplets. First, we consider even elements of the Poincar\'e superalgebra  \rf{11052019-man-01}. Realizations of the Poincar\'e algebra \rf{11052019-man-01} in terms of differential operators acting on the fields $\phi_\lambda(p)$ and $\psi_\lambda(p)$ \rf{15052019-man02-01}-\rf{15052019-man02-06} is given by the well known expressions.

\vspace{-0.7cm}
\beq
&& \hbox{ \it Realizations on space of $\phi_\lambda(p)$ and $\psi_\lambda(p)$}:
\nonumber\\[-7pt]
\label{15052019-man02-12} && P^\Rsm = p^\Rsm\,,  \qquad P^\Lsm = p^\Lsm\,,   \qquad   \hspace{0.8cm} P^+=\beta\,,\qquad
P^- = p^-\,, \qquad p^- \equiv - \frac{p^\Rsm p^\Lsm}{\beta}\,,\qquad
\\
\label{15052019-man02-13} && J^{+\Rsm}= \irm x^+ P^\Rsm + \partial_{p^\Lsm}\beta\,, \hspace{2cm} J^{+\Lsm}= \irm x^+ P^\Lsm + \partial_{p^\Rsm}\beta\,, \
\\
 \label{15052019-man02-15} && J^{+-} = \irm x^+P^- + \partial_\beta \beta - \half e_\lambda\,,  \hspace{1cm} J^{\Rsm\Lsm} =  p^\Rsm\partial_{p^\Rsm} - p^\Lsm\partial_{p^\Lsm} + \lambda\,,
\\
\label{15052019-man02-17} && J^{-\Rsm} = -\partial_\beta p^\Rsm + \partial_{p^\Lsm} p^- + \lambda \frac{p^\Rsm}{\beta} + \frac{p^\Rsm}{2\beta}e_\lambda\,,
\\
\label{15052019-man02-18} && J^{-\Lsm} = -\partial_\beta p^\Lsm + \partial_{p^\Rsm} p^-
- \lambda\frac{p^\Lsm}{\beta} + \frac{p^\Lsm}{2\beta} e_\lambda\,,
\eeq
where the notation for partial derivatives and the definition of symbol $e_\lambda$ are given by
\beq
\label{15052019-man02-19} && \partial_\beta\equiv \partial/\partial \beta\,, \quad \partial_{p^\Rsm}\equiv \partial/\partial p^\Rsm\,, \qquad \partial_{p^\Lsm}\equiv \partial/\partial p^\Lsm\,,
\\
\label{15052019-man02-20} && e_\lambda =0 \hspace{0.5cm} \hbox{ for integer } \lambda\,,\hspace{1.4cm}
e_\lambda = 1 \hspace{0.5cm} \hbox{ for half-integer } \ \lambda\,,
\\
\label{15052019-man02-21} && e_\lambda + e_{\lambda+\half} = 1\,, \qquad e_\lambda e_{\lambda+\half} = 0\,.
\eeq
In \rf{15052019-man02-21}, we present relations which follow from the definition of $e_\lambda$ given in \rf{15052019-man02-20}.

Having presented realization of the Poincar\'e algebra
in terms of  differential operators in \rf{15052019-man02-12}-\rf{15052019-man02-18} we are able to provide a field representation for generators in \rf{11052019-man-01}. To quadratic order in fields, a field representation of the Poincar\'e algebra generators \rf{11052019-man-01} is given by
\beq
\label{15052019-man02-22} && \hspace{-4cm} G_\smpt  =  \sum_{s=1}^\infty G_\smpt^{(s)} +  \sum_{s=0}^\infty G_\smpt^{(s+\half)}\,,
\\
\label{15052019-man02-23} G_\smpt^{(s)} & = & 2 \int d^3p\,\, \big( \beta \phi_s^\dagger G_\diff \phi_s + \phi_{s-\half}^\dagger G_\diff \phi_{s-\half}\big)\,,
\\
\label{15052019-man02-24} G_\smpt^{(s+\half)} & = & 2 \int d^3p\,\, \big( \beta \psi_s^\dagger G_\diff \psi_s + \psi_{s+\half}^\dagger G_\diff \psi_{s+\half}\big)\,,
\eeq
where $G_\diff$ denotes the differential operators presented in
\rf{15052019-man02-12}-\rf{15052019-man02-18}, while
$G_\smpt$ denotes the field representation for the generators \rf{11052019-man-01}. For the odd elements of the Poincar\'e superalgebra  (supercharges $Q^{\pm \Rsm,\Lsm}$), a field representation $G_\smpt$ takes the form  as in \rf{15052019-man02-22}, where
\beq
\label{15052019-man02-25} && Q_\smpt^{+\Rsm\, (s)} = 2 \int d^3p\,\, \beta  \phi_s^\dagger \phi_{s-\half} \,, \hspace{1.9cm} Q_\smpt^{+\Lsm\, (s)} = 2 \int d^3p\,\, \beta   \phi_{s-\half}^\dagger \phi_s \,,
\\
\label{15052019-man02-26} && Q_\smpt^{-\Rsm\, (s)} = 2 \int d^3p\,\, p^\Rsm  \phi_{s-\half}^\dagger  \phi_s \,, \hspace{1.8cm} Q_\smpt^{-\Lsm\, (s)} = 2 \int d^3p\,\, p^\Lsm \phi_s^\dagger  \phi_{s-\half}  \,,
\\
\label{15052019-man02-27} && Q_\smpt^{+\Rsm\, (s+\half)} = - 2 \int d^3p\,\, \beta  \psi_{s+\half}^\dagger \psi_s  \,,
\hspace{1.1cm} Q_\smpt^{+\Lsm\, (s+\half)} = - 2 \int d^3p\,\, \beta  \psi_s^\dagger \psi_{s+\half} \,,
\\
\label{15052019-man02-28} && Q_\smpt^{-\Rsm\, (s+\half)} = - 2 \int d^3p\,\, p^\Rsm   \psi_s^\dagger  \psi_{s+\half}   \,, \hspace{1cm} Q_\smpt^{-\Lsm\, (s+\half)} = - 2 \int d^3p\,\, p^\Lsm \psi_{s+\half}^\dagger  \psi_s \,.\qquad
\eeq

The fields $\phi_\lambda$, $\psi_\lambda$ satisfy the Poisson-Dirac equal-time commutation relations
{\small
\beq
\label{15052019-man02-29} && \hspace{-2cm} [\phi_\lambda(p),\phi_{\lambda'}^\dagger(p')] = \frac{\delta_{\lambda\lambda'}}{2\beta}\delta^3(p-p') \,,  \hspace{0.6cm} [\psi_\lambda(p),\psi_{\lambda'}^\dagger(p')] = \frac{\delta_{\lambda\lambda'}}{2\beta}\delta^3(p-p')\,, \quad \hbox{ integer} \ \lambda,\lambda' \,,
\\
\label{15052019-man02-30} &&\hspace{-2cm} \{\phi_\lambda(p),\phi_{\lambda'}^\dagger(p')\} = \frac{\delta_{\lambda\lambda'}}{2}\delta^3(p-p')\,, \quad \{\psi_\lambda(p),\psi_{\lambda'}^\dagger(p')\} = \frac{\delta_{\lambda\lambda'}}{2}\delta^3(p-p')\,, \quad \hbox{half-integer } \lambda,\lambda'\,.
\eeq
}
Using relations given in \rf{15052019-man02-22}-\rf{15052019-man02-24}, we verify the standard equal-time commutation relations between the fields and the even generators
\be \label{15052019-man02-31}
[ \phi_\lambda,G_\smpt\,] =  G_\diff \phi_\lambda \,, \qquad [\psi_\lambda,G_\smpt\,] =  G_\diff \psi_\lambda \,,
\ee
while using expressions in \rf{15052019-man02-25}-\rf{15052019-man02-28}, we find the following equal-time (anti)commutation relations between the supercharges and the fields
\beq
\label{15052019-man02-32} && [\phi_s,Q^{+\Rsm}] = \phi_{s-\half}\,, \hspace{2.2cm} \{ \phi_{s-\half} , Q^{+\Lsm}\} = \beta \phi_s\,,
\\
\label{15052019-man02-33} && \{\phi_{s-\half},Q^{-\Rsm}\} = p^\Rsm\phi_s\,, \hspace{1.7cm}  [ \phi_s , Q^{-\Lsm}] = \frac{p^\Lsm}{\beta} \phi_{s-\half}\,,
\\
\label{15052019-man02-34} && \{ \psi_{s+\half} , Q^{+\Rsm}\} = - \beta \psi_s\,,  \hspace{1.5cm} [\psi_s,Q^{+\Lsm}] =  - \psi_{s+\half}\,,
\\
\label{15052019-man02-35} && [\psi_s , Q^{-\Rsm}] = - \frac{p^\Rsm}{\beta} \psi_{s+\half}\,,
\hspace{1.5cm}  \{\psi_{s+\half},Q^{-\Lsm}\} =  - p^\Lsm\psi_s\,.
\eeq

\newsection{ \large   Superfield formulation of free massless $N=1$ supermultiplets}   \label{sec-superfield}

In order to discuss a light-cone gauge superfield formulation we introduce a Grassmann-odd momentum denoted by $p_\theta$. The momentum superspace is parametrized by the light-cone time $x^+$, the momenta $p^\Rsm$, $p^\Lsm$, $\beta$ and the Grassmann momentum $p_\theta$,
\be \label{16052019-man02-01-x}
x^+\,, \beta\,, \ p^\Rsm\,, \  p^\Lsm\,, \ p_\theta\,.
\ee
Using component fields \rf{15052019-man02-04}-\rf{15052019-man02-07} depending on $x^+$ and momenta $p^\Rsm$, $p^\Lsm$, $\beta$ , we introduce then unconstrained superfields $\Phi_s$, $\Phi_{-s+\half}$ and $\Psi_{s+\half}$, $\Psi_{-s}$ defined in the superspace \rf{16052019-man02-01-x} in the following way:
\beq
\label{16052019-man02-01} && \hspace{-1.5cm} \Phi_s = \phi_s + \frac{p_\theta}{\beta} \phi_{s-\half}\,, \hspace{0.8cm}  \Phi_{-s+\half} = \phi_{-s+\half} + p_\theta \phi_{-s}\,, \hspace{0.5cm} \hbox{ for spin-$s$ supermultiplets,}
\\
\label{16052019-man02-02} && \hspace{-1.5cm} \Psi_{s+\half} = p_\theta \psi_s +  \psi_{s+\half}\,, \hspace{0.4cm} \Psi_{-s} = \psi_{-s} + \frac{p_\theta}{\beta} \psi_{-s-\half}\,, \hspace{0.8cm} \hbox{ for spin-$(s+\half)$ supermultiplets,}
\\
\label{16052019-man02-03} && \hspace{-1.5cm} \Psi_\half = p_\theta \psi_0 +  \psi_\half\,, \hspace{1.2cm} \Psi_0 = \psi_{-0} + \frac{p_\theta}{\beta} \psi_{-\half}\,, \hspace{1.2cm} \hbox{ for spin-$\half$ supermultiplets,}
\eeq
where, in \rf{16052019-man02-01},\rf{16052019-man02-02}, $s=1,\ldots,\infty$.

Our basic observation which considerably simplifies our analysis of theory of interacting superfields is that the unconstrained superfields \rf{16052019-man02-01}-\rf{16052019-man02-03} can be collected into unconstrained superfields denoted as $\Theta_\lambda$,
\be \label{16052019-man02-04}
\Theta_\lambda(p,p_\theta)\,, \qquad \lambda = \hbox{$0,\pm \half,\pm 1,\ldots ,\pm \infty$}\,,
\ee
where, depending on $\lambda$, the superfield  $\Theta_\lambda$ is identified with the ones in \rf{16052019-man02-01}-\rf{16052019-man02-03}  as follows:
\beq
&& \Theta_s \equiv \Phi_s\,, \hspace{2.8cm} \Theta_{-s+\half} \equiv \Phi_{-s+\half}\,, \hspace{1cm} s =1,2,\ldots, \infty\,,
\nonumber\\
\label{16052019-man02-07} && \Theta_{s+\half} \equiv \Psi_{s+\half}\,, \hspace{2cm} \Theta_{-s} \equiv \Psi_{-s}\,, \hspace{1.8cm} s =0,1,\ldots, \infty\,.
\eeq
We note the following property of the superfield $\Theta_\lambda$. Using the notation $\GP(\Theta_\lambda)$ for the Grassmann parity of the superfield $\Theta_\lambda$ and taking into account definition of $e_\lambda$ \rf{15052019-man02-20}, we note the relation,
\be \label{16052019-man02-07-a1}
\GP(\Theta_\lambda) = {e_\lambda}\,.
\ee
We see that, for integer $\lambda$, the superfield $\Theta_\lambda$ is Grassmann even, while, for half-integer $\lambda$, the superfield $\Theta_\lambda$ is Grassmann odd.

\noindent {\bf Realizations of the Poincar\'e superalgebra on superfield $\Theta_\lambda$}. Realization of the Poincar\'e superalgebra in terms of differential operators acting on the superfield $\Theta_\lambda(p,p_\theta)$ is given by
\beq
\label{16052019-man02-10} && P^\Rsm = p^\Rsm\,,  \qquad P^\Lsm = p^\Lsm\,,   \hspace{2cm}   P^+=\beta\,,\qquad
P^- = p^-\,, \qquad p^-\equiv - \frac{p^\Rsm p^\Lsm}{\beta}\,,\qquad
\\
\label{16052019-man02-11} && J^{+\Rsm}= \irm x^+ P^\Rsm + \partial_{p^\Lsm}\beta\,,
\hspace{2.4cm}  J^{+\Lsm}= \irm x^+ P^\Lsm + \partial_{p^\Rsm}\beta\,, \
\\
\label{16052019-man02-12} && J^{+-} = \irm x^+P^- + \partial_\beta \beta + M_\lambda^{+-}\,, \hspace{1cm} J^{\Rsm\Lsm} =  p^\Rsm\partial_{p^\Rsm} - p^\Lsm\partial_{p^\Lsm} + M_\lambda^{\Rsm\Lsm}\,,
\\
\label{16052019-man02-14} && J^{-\Rsm} = -\partial_\beta p^\Rsm + \partial_{p^\Lsm} p^-
+ M_\lambda^{\Rsm\Lsm}\frac{p^\Rsm}{\beta} - M_\lambda^{+-} \frac{p^\Rsm}{\beta}\,,
\\
\label{16052019-man02-15} && J^{-\Lsm} = -\partial_\beta p^\Lsm + \partial_{p^\Rsm} p^-
- M_\lambda^{\Rsm\Lsm}\frac{p^\Lsm}{\beta} - M_\lambda^{+-} \frac{p^\Lsm}{\beta}\,,
\\
\label{16052019-man02-16} && \hspace{1.2cm} M_\lambda^{+-} =  \half p_\theta\partial_{p_\theta} - \half e_\lambda\,, \hspace{2cm} M_\lambda^{\Rsm\Lsm}  = \lambda -\half p_\theta\partial_{p_\theta}
\\
\label{16052019-man02-17-a}&& Q^{+\Rsm} = (-)^{e_\lambda} \beta \partial_{p_\theta}\,,
\hspace{2.3cm}  Q^{+\Lsm} = (-)^{e_\lambda} p_\theta\,,
\\
\label{16052019-man02-17} && Q^{-\Rsm} =  (-)^{e_\lambda} \frac{1}{\beta} p^\Rsm p_\theta\,, \hspace{2cm} Q^{-\Lsm} = (-)^{e_\lambda} p^\Lsm \partial_{p_\theta}\,,
\eeq
where the symbol $e_\lambda$ is defined in \rf{15052019-man02-20}, while a quantity $\partial_{p_\theta}$ stands for left derivative w.r.t the Grassmann momenta $p_\theta$ (see Appendix A). Explicit realization of the Poincar\'e superalgebra on the superfields $\Phi$ and $\Psi$ \rf{16052019-man02-01}-\rf{16052019-man02-03} is given in Table I.

\medskip
\noindent{\small\sf Table I. Realization of supercharges \rf{16052019-man02-17-a},\rf{16052019-man02-17} on superfields \rf{16052019-man02-01}-\rf{16052019-man02-03}. Realization of the Poincar\'e algebra on superfields \rf{16052019-man02-01}-\rf{16052019-man02-03} is given by relations \rf{16052019-man02-10}-\rf{16052019-man02-15}, where the operators $M_\lambda^{+-}$, $M_\lambda^{\Rsm\Lsm}$ should be replaced by operators $M^{+-}$, $M^{\Rsm\Lsm}$ given in this Table.}
{\small
\begin{center}
\begin{tabular}{|c|c|c|c|c|}
\hline
& & & &
\\[-3mm]
& $\Phi_s$ & $\Phi_{-s+\half}$  & $\Psi_{s+\half}$ & $\Psi_{-s}$
\\[2mm]
\hline
&&&&
\\[-3mm]
$Q^{+\Rsm}$ & $\beta\partial_{p_\theta}$ & $-\beta\partial_{p_\theta}$  & $-\beta\partial_{p_\theta}$ & $\beta\partial_{p_\theta}$
\\[2mm]
\hline
&&&&
\\[-3mm]
$Q^{+\Lsm}$ & $p_\theta$ & $-p_\theta$  & $-p_\theta$ & $p_\theta$
\\[2mm]
\hline
&&&&
\\[-3mm]
$Q^{-\Rsm}$ & $\frac{1}{\beta}p^\Rsm p_\theta$ & $-\frac{1}{\beta}p^\Rsm p_\theta$  & $-\frac{1}{\beta}p^\Rsm p_\theta$ & $\frac{1}{\beta}p^\Rsm p_\theta$
\\[2mm]
\hline
&&&&
\\[-3mm]
$Q^{-\Lsm}$ & $p^\Lsm\partial_{p_\theta}$ & $-p^\Lsm\partial_{p_\theta}$  & $-p^\Lsm\partial_{p_\theta}$ & $p^\Lsm\partial_{p_\theta}$
\\[2mm]
\hline
&&&&
\\[-3mm]
$M^{+-}$ & $ \half p_\theta\partial_{p_\theta}$ & $\half p_\theta\partial_{p_\theta}-\half$  & $\half p_\theta\partial_{p_\theta}-\half$ & $\half p_\theta\partial_{p_\theta}$
\\[2mm]
\hline
&&&&
\\[-3mm]
$M^{\Rsm\Lsm}$ & $ s-\half p_\theta\partial_{p_\theta}$ & $-s-\half p_\theta\partial_{p_\theta}+\half$  & $s-\half p_\theta\partial_{p_\theta}+\half$ & $-s-\half p_\theta\partial_{p_\theta}$
\\[2mm]
\hline
\end{tabular}
\end{center}
}

\medskip
In addition to the superfields $\Theta_\lambda$, we find it convenient to use other superfields denoted by $\Theta_\lambda^*$. The superfields $\Theta_\lambda^*$ are constructed out of the hermitian conjugated fields $\phi_\lambda^\dagger$, $\psi_\lambda^\dagger$ and defined as follows. First, we define superfields $\Phi_\lambda^*$, $\Psi_\lambda^*$ by the relations

\newpage

\beq
\label{16052019-man02-18} && \hspace{-1.5cm} \Phi_{s-\half}^* =  p_\theta \phi_s^\dagger - \phi_{s-\half}^\dagger\,, \hspace{0.6cm}  \Phi_{-s}^* = \phi_{-s}^\dagger + \frac{p_\theta}{\beta} \phi_{-s+\half}^\dagger\,, \hspace{0.8cm} \hbox{ for spin-$s$ supermultiplets};
\\
\label{16052019-man02-19} && \hspace{-1.5cm} \Psi_s^* =  \psi_s^\dagger +  \frac{p_\theta}{\beta} \psi_{s+\half}^\dagger\,, \hspace{0.9cm} \Psi_{-s-\half}^* = p_\theta \psi_{-s}^\dagger -  \psi_{-s-\half}^\dagger\,, \hspace{0.4cm} \hbox{ for spin-$(s+\half)$ supermultiplets};\qquad
\\
\label{16052019-man02-20} && \hspace{-1.5cm} \Psi_0^* =  \psi_0^\dagger +  \frac{p_\theta}{\beta} \psi_\half^\dagger\,, \hspace{1.3cm} \Psi_{-\half}^* = p_\theta \psi_{-0}^\dagger -  \psi_{-\half}^\dagger\,, \hspace{1.1cm} \hbox{ for spin-$\half$ supermultiplets};\qquad
\eeq
where, in \rf{16052019-man02-18},\rf{16052019-man02-19},  $s=1,\ldots, \infty$.  Second, we introduce superfields $\Theta_\lambda^*$ defined for all $\lambda$,
\be \label{16052019-man02-21}
\Theta_\lambda^*(p,p_\theta)\,, \qquad  \lambda = \hbox{$ 0,\pm \half,\pm 1,\ldots ,\pm \infty$};
\ee
where, depending on $\lambda$, the superfield  $\Theta_\lambda^*$ is identified with the ones in \rf{16052019-man02-18}-\rf{16052019-man02-20} as
\beq
\label{16052019-man02-22} && \Theta_{s-\half}^* \equiv \Phi_{s-\half}^*\,, \hspace{1.3cm} \Theta_{-s}^* \equiv \Phi_{-s}^*\,,   \hspace{1.8cm} s=1,2,\ldots, \infty;
\nonumber\\
&& \Theta_s^* \equiv \Psi_s^*\,, \hspace{2.1cm} \Theta_{-s-\half}^* \equiv \Psi_{-s-\half}^*\,,  \hspace{1cm} s=0,1,\ldots, \infty\,.
\eeq

The new superfields $\Theta_\lambda^*$ are not independent of the superfields $\Theta_\lambda$. Namely, in view of the hermitian conjugation rule given in \rf{15052019-man02-11}, we find the relation
\be \label{16052019-man02-23}
\Theta_{-\lambda}^*(-p,-p_\theta) =  (-)^{e_\lambda} \Theta_\lambda(p,p_\theta)\,,
\ee
where $e_{\lambda}$ is defined in \rf{15052019-man02-20} and we show explicitly momentum arguments $p$, $p_\theta$ entering the superfields. For integer $\lambda$, the superfield $\Theta_\lambda^*$ is Grassmann even, while, for half-integer $\lambda$, the superfield $\Theta_\lambda^*$ is Grassmann odd. In other words, the Grassmann parity of the superfield $\Theta_\lambda^*$, is given by the relation $\GP(\Theta_\lambda^*) = e_\lambda$.

Using the realization of the Poincar\'e superalgebra
in terms of  differential operators in \rf{16052019-man02-10}-\rf{16052019-man02-17}, we can present  a superfield representation for generators in \rf{11052019-man-05},\rf{11052019-man-06}. To quadratic order in the superfields $\Theta_\lambda$, a superfield representation of Poincar\'e superalgebra generators \rf{11052019-man-05}, \rf{11052019-man-06} is given by
\be
\label{16052019-man02-24}  G_\smpt  =  \sum_{\lambda=-\infty}^{+\infty} G_{\smpt,\,\lambda}  \qquad G_{\smpt,\, \lambda}  = \int d^3p dp_\theta \,\, \beta \Theta_{\lambda-\half}^* G_{\diff,\,\lambda }\Theta_\lambda\,,
\ee
where realization of $G_{\diff,\, \lambda}$ on space of $\Theta_\lambda$ is given in \rf{16052019-man02-10}-\rf{16052019-man02-17}.
A realization of $G_{\diff,\, \lambda}$ on space of the superfield $\Theta_\lambda^*$ may be found in Appendix B.

The superfields $\Theta_\lambda$, $\Theta_\lambda^*$ satisfy the Poisson-Dirac equal-time commutation relations
\be \label{16052019-man02-26}
[\Theta_\lambda(p,p_\theta),\Theta_{\lambda'}^*(p',p_\theta')]_\pm = \frac{(-)^{ e_{\lambda+\half} } }{2\beta}\delta^3(p-p')\delta(p_\theta-p_\theta') \delta_{\lambda-\lambda',\half}\,,
\ee
where $[a,b]_\pm$ stands for a graded commutator, $[a,b]_\pm = (-)^{\epsilon_a \epsilon_b +1} [b,a]_\pm$. Using relations given in \rf{16052019-man02-24},\rf{16052019-man02-26}, we verify the standard equal-time (anti)commutation relation between the superfields and the generators
\be \label{16052019-man02-27}
[\Theta_\lambda,G_\smpt]_{\pm} =  G_{\diff,\,\lambda} \Theta_\lambda \,,
\ee
where $G_{\diff,\,\lambda}$ are given in \rf{16052019-man02-10}-\rf{16052019-man02-17}.

In light-cone gauge Lagrangian approach, the light-cone gauge action takes the form
\be  \label{27062018-man02-31-a1}
S = \half\sum_{\lambda=-\infty}^\infty \int dx^+ d^3p dp_\theta \,\, \Theta_{\lambda-\half}^* \big( 2\irm \beta \partial^- - 2p^\Rsm p^\Lsm \big)\Theta_\lambda  +\int dx^+ P_{\rm int}^-\,,
\ee
where $\partial^-\equiv\partial/\partial x^+$ and $P_{\rm int}^-$ is a light-cone gauge Hamiltonian describing interactions. Internal symmetry can be incorporated via the Chan--Paton method used in string theory (see Sec.6).

\newsection{ \large General structure of $n$-point dynamical generators of the Poincar\'e superalgebra} \label{sec-npoint}

We now describe a general structure of the dynamical generators of the Poincar\'e
superalgebra. For theories of interacting fields, the Poincar\'e superalgebra dynamical generators receive corrections having higher powers of fields. In general,  one has the following
expansion for the dynamical generators
\be \label{14052019-man02-01}
G^\dyn
= \sum_{n=2}^\infty
G_\smpn^\dyn\,,
\ee
where $G_\smpn^\dyn$ in \rf{14052019-man02-01} stands for a functional
that has $n$ powers of superfields $\Theta^*$.

In this Section, for arbitrary $n\geq 3$, we describe restrictions imposed on the dynamical generators $G_\smpn^\dyn$ by the kinematical symmetries of the Poincar\'e superalgebra. We discuss the restrictions in turn.

\noindent {\bf Kinematical $P^{\Rsm,\Lsm}$, $P^+$, $Q^{+\Lsm}$ symmetries.}. Using (anti)commutation relations between the dynamical generators given in \rf{11052019-man-06} and the kinematical generators $P^\Rsm$, $P^\Lsm$, $P^+$, $Q^{+\Lsm}$,  we verify that the dynamical generators $G_\smpn^\dyn$ with $n\geq 3$ take the following form:
\beq
\label{14052019-man02-02} && P_\smpn^- = \int\!\! d\Gamma_\smpn\,\,  \langle \Theta_\smpn^*  | p_\smpn^-\rangle\,,
\\
\label{14052019-man02-03} && Q_\smpn^{-\Rsm} = \int\!\! d\Gamma_\smpn\,\,  \langle \Theta_\smpn^* | q_\smpn^{-\Rsm}\rangle\,,
\\
\label{14052019-man02-04} && Q_\smpn^{-\Lsm} = \int\!\! d\Gamma_\smpn\,\,  \langle \Theta_\smpn^*  | q_\smpn^{-\Lsm} \rangle\,,
\\
\label{14052019-man02-05} && J_\smpn^{-\Rsm} = \int\!\! d\Gamma_\smpn\,\,  \langle\Theta_\smpn^* | j_\smpn^{-\Rsm}\rangle  +   \langle \Xbf_\smpn^\Rsm \Theta_\smpn^* | p_\smpn^-\rangle  + (-)^n \langle \Xbf_{\smpn\,\theta } \Theta_\smpn^* |q_\smpn^{-\Rsm} \rangle\,,
\\
\label{14052019-man02-06} && J_\smpn^{-\Lsm} = \int\!\! d\Gamma_\smpn\,\,  \langle \Theta_\smpn^* | j_\smpn^{-\Lsm}\rangle +   \langle \Xbf_\smpn^\Lsm \Theta_\smpn^* | p_\smpn^- \rangle  + \frac{(-)^{n+1}}{n} \PP_{\smpn\,\theta } \langle \Theta_\smpn^*   | q_\smpn^{-\Lsm} \rangle\,,
\eeq
where we use the notation
\beq
\label{14052019-man02-07} && d\Gamma_\smpn = d\Gamma_\smpn^p d\Gamma_\smpn^{p_\theta} \,,
\\
\label{14052019-man02-08} && d\Gamma_\smpn^p =  \delta^{3}(\sum_{a=1}^n p_a)\prod_{a=1}^n \frac{d^3p_a}{(2\pi)^{3/2} }\,, \qquad d^3 p_a = dp_a^\Rsm dp_a^\Lsm d\beta_a\,,
\\
\label{14052019-man02-09} && d\Gamma_\smpn^{p_\theta} \equiv dp_{\theta_1} \ldots dp_{\theta_n} \delta(\sum_{a=1}^n  p_{\theta_a} )\,,
\\
\label{14052019-man02-10} && \Xbf_\smpn^\Rsm =  - \frac{1}{n}\sum_{a=1}^n \partial_{p_a^\Lsm}\,, \hspace{1cm} \Xbf_\smpn^\Lsm = - \frac{1}{n}\sum_{a=1}^n\partial_{p_a^\Rsm}\,,
\\
\label{14052019-man02-11} && \Xbf_{\smpn\,\theta} = \frac{1}{n}\sum_{a=1}^n \partial_{p_{\theta_a}}\,,\hspace{1cm} \PP_{\smpn \theta}=  \sum_{a=1}^n \frac{p_{\theta_a}}{\beta_a}\,.
\eeq
In \rf{14052019-man02-02}-\rf{14052019-man02-06}, expressions $\langle \Theta_\smpn| p_\smpn^-\rangle$, $\langle \Theta_\smpn| q_\smpn^{-\Rsm,\Lsm}\rangle$, and $\langle \Theta_\smpn| j_\smpn^{-\Rsm,\Lsm}\rangle$ stand for shortcuts defined  as
\beq
\label{14052019-man02-12} && \langle \Theta_\smpn| p_\smpn^-\rangle \quad \equiv \sum_{\lambda_1\ldots\lambda_n} \Theta_{\lambda_1\ldots\lambda_n}^*  p_{\lambda_1\ldots\lambda_n}^-\,,
\\
\label{14052019-man02-13} && \langle \Theta_\smpn| q_\smpn^{-\Rsm,\Lsm} \rangle \equiv \sum_{\lambda_1\ldots\lambda_n} \Theta_{\lambda_1\ldots\lambda_n}^*  q_{\lambda_1\ldots\lambda_n}^{-\Rsm,\Lsm}\,,
\\
\label{14052019-man02-14} && \langle \Theta_\smpn| j_\smpn^{-\Rsm,\Lsm} \rangle \equiv \sum_{\lambda_1\ldots\lambda_n} \Theta_{\lambda_1\ldots\lambda_n}^*  j_{\lambda_1\ldots\lambda_n}^{-\Rsm,\Lsm}\,,
\\
\label{14052019-man02-15} && \hspace{2cm} \Theta_{\lambda_1\ldots\lambda_n}^* \equiv  \Theta_{\lambda_1}^*(p_1,p_{\theta_1})  \ldots  \Theta_{\lambda_n}^*(p_n,p_{\theta_n}) \,.
\eeq
To simplify our presentation, the quantities $p_{\lambda_1\ldots\lambda_n}^-$, $q_{\lambda_1\ldots\lambda_n}^{-\Rsm,\Lsm}$, and $j_{\lambda_1\ldots\lambda_n}^{-\Rsm,\Lsm}$ appearing in \rf{14052019-man02-12}-\rf{14052019-man02-14}, will shortly be denoted as $g_{\lambda_1\ldots\lambda_n}$,
\be \label{14052019-man02-16}
g_{\lambda_1\ldots\lambda_n} = p_{\lambda_1\ldots\lambda_n}^-,\quad
q_{\lambda_1\ldots\lambda_n}^{-\Rsm},\quad
q_{\lambda_1\ldots\lambda_n}^{-\Lsm},\quad  j_{\lambda_1\ldots\lambda_n}^{-\Rsm},\quad
j_{\lambda_1\ldots\lambda_n}^{-\Lsm}\,.
\ee
We refer to quantities $g_{\lambda_1\ldots\lambda_n}$ \rf{14052019-man02-16} as $n$-point densities. We note that $n$-point densities $g_{\lambda_1\ldots\lambda_n}$ \rf{14052019-man02-16} depend on the momenta $p_a^\Rsm$, $p_a^\Lsm$, $\beta_a$, Grassmann momenta $p_{\theta_a}$, and helicities $\lambda_a$, $a=1,2\ldots,n$,
\be \label{14052019-man02-17}
g_{\lambda_1\ldots\lambda_n} = g_{\lambda_1\ldots\lambda_n}(p_a,p_{\theta_a})\,.
\ee
Note that we use the indices $a,b=1,\ldots,n$ to label superfields entering $n$-point interaction vertex. Also note that, in \rf{14052019-man02-02}-\rf{14052019-man02-06}, the differential operators $\Xbf_\smpn^{\Rsm,\Lsm}$, $\Xbf_{\smpn\,\theta}$ are acting only on the arguments of the superfields. For example, the expression $\langle \Xbf^\Rsm \Theta_\smpn^* | g_\smpn \rangle$ should read as
\be \label{14052019-man02-17-a}
\langle \Xbf^\Rsm \Theta_\smpn^* | g_\smpn\rangle =  \sum_{\lambda_1,\ldots \lambda_n} (\Xbf^\Rsm\Theta_{\lambda_1\ldots\lambda_n}^* ) g_{\lambda_1\ldots\lambda_n}\,.
\ee
Note that the argument $p_a$ in \rf{14052019-man02-08} stands for the momenta $p_a^\Rsm$, $p_a^\Lsm$, and $\beta_a$.
In what follows, the density $p_\smpn^-$ will often be referred to as an $n$-point interaction vertex, while, for $n=3$, we refer to density $p_\smp3^-$ as cubic interaction vertex.

\noindent {\bf $J^{+-}$-symmetry equations}. Commutation relations between the dynamical generators $P^-$, $Q^{-\Rsm,\Lsm}$, $J^{-\Rsm,\Lsm}$ and the kinematical generator $J^{+-}$ amount to equations for the densities given by:
\beq
\label{14052019-man02-18} && \hspace{-1.5cm} \sum_{a=1}^n  \big( \beta_a\partial_{\beta_a} + \half p_{\theta_a}\partial_{p_{\theta_a}} + \half e_{\lambda_a} \big) g_{\lambda_1\ldots \lambda_n} =   \frac{n-1}{2} g_{\lambda_1\ldots \lambda_n}\,, \ \ \hbox{ for }\ g_{\lambda_1\ldots \lambda_n} = p_{\lambda_1\ldots \lambda_n}^-\,, j_{\lambda_1\ldots \lambda_n}^{-\Rsm,\Lsm}\,,\quad
\\
\label{14052019-man02-19} && \hspace{-1.5cm} \sum_{a=1}^n  \big( \beta_a\partial_{\beta_a} + \half p_{\theta_a}\partial_{p_{\theta_a}} + \half e_{\lambda_a} \big) g_{\lambda_1\ldots \lambda_n} =  \frac{n}{2}  g_{\lambda_1\ldots \lambda_n}\,, \hspace{1cm} \hbox{ for }\ g_{\lambda_1\ldots \lambda_n} = q_{\lambda_1\ldots \lambda_n}^{-\Rsm,\Lsm}\,.\quad
\eeq

\noindent {\bf $J^{\Rsm\Lsm}$-symmetry equations}. Commutation relations between the dynamical generators $P^-$, $Q^{-\Rsm,\Lsm}$, $J^{-\Rsm,\Lsm}$ and the kinematical generator $J^{\Rsm\Lsm}$ amount to equations for the densities given by
\beq
\label{14052019-man02-20} && \sum_{a=1}^n  \big( p_a^\Rsm\partial_{p_a^\Rsm} - p_a^\Lsm\partial_{p_a^\Lsm} - \half p_{\theta_a} \partial_{p_{\theta_a}} + \lambda_a \big) p_{\lambda_1\ldots \lambda_n} = - \frac{n-1}{2} p_{\lambda_1\ldots \lambda_n}^-\,,
\\
\label{14052019-man02-21} && \sum_{a=1}^n  \big( p_a^\Rsm\partial_{p_a^\Rsm} - p_a^\Lsm\partial_{p_a^\Lsm} - \half p_{\theta_a} \partial_{p_{\theta_a}} + \lambda_a \big) q_{\lambda_1\ldots \lambda_n}^{-\Rsm} =  - \frac{n-2}{2}   q_{\lambda_1\ldots \lambda_n}^{-\Rsm}\,,
\\
\label{14052019-man02-22} && \sum_{a=1}^n  \big( p_a^\Rsm\partial_{p_a^\Rsm} - p_a^\Lsm\partial_{p_a^\Lsm} - \half p_{\theta_a} \partial_{p_{\theta_a}} + \lambda_a \big) q_{\lambda_1\ldots \lambda_n}^{-\Lsm} =  - \frac{n}{2}   q_{\lambda_1\ldots \lambda_n}^{-\Lsm}\,,
\\
\label{14052019-man02-23} && \sum_{a=1}^n  \big( p_a^\Rsm\partial_{p_a^\Rsm} - p_a^\Lsm\partial_{p_a^\Lsm} - \half p_{\theta_a} \partial_{p_{\theta_a}} + \lambda_a \big) j_{\lambda_1\ldots \lambda_n}^{-\Rsm} = - \frac{n-3}{2}   j_{\lambda_1\ldots \lambda_n}^{-\Rsm}\,,
\\
\label{14052019-man02-24} && \sum_{a=1}^n  \big( p_a^\Rsm\partial_{p_a^\Rsm} - p_a^\Lsm\partial_{p_a^\Lsm} - \half p_{\theta_a} \partial_{p_{\theta_a}} + \lambda_a \big) j_{\lambda_1\ldots \lambda_n}^{-\Lsm} =  - \frac{n+1}{2}j_{\lambda_1\ldots \lambda_n}^{-\Lsm}\,.
\eeq

\noindent {\bf $J^{+\Rsm}$, $J^{+\Lsm}$, $Q^{+\Rsm}$-symmetry equations}. Using (anti)commutation relations between the dynamical generators $P^-$, $Q^{-\Rsm,\Lsm}$, $J^{-\Rsm,\Lsm}$ and the kinematical generators $J^{+\Rsm}$, $J^{+\Lsm}$, and $Q^{+\Rsm}$, we find that the densities $g_{\lambda_1\ldots\lambda_n}$ \rf{14052019-man02-16} depend on the momenta $p_a^{\Rsm,\Lsm}$ and the Grassmann momenta $p_{\theta_a}$ through new momentum variables $\Po_{ab}^{\Rsm,\Lsm}$ and $\Po_{\theta\, ab}$  defined by the relations
\be \label{14052019-man02-25}
\Po_{ab}^\Rsm \equiv p_a^\Rsm \beta_b - p_b^\Rsm \beta_a\,, \qquad
\Po_{ab}^\Lsm \equiv p_a^\Lsm \beta_b - p_b^\Lsm \beta_a\,, \qquad
\Po_{\theta\, ab} \equiv p_{\theta_a} \beta_b - p_{\theta_b} \beta_a\,.
\ee
This is to say that our densities $g_{\lambda_1\ldots\lambda_n}$ \rf{14052019-man02-16} turn out to be functions of $\Po_{ab}^{\Rsm,\Lsm}$ and $\Po_{\theta\, ab}$ in place of $p_a^{\Rsm,\Lsm}$,
$p_{\theta_a}$,
\beq
&& g_{\lambda_1\ldots\lambda_n} = g_{\lambda_1\ldots\lambda_n} (\Po_{ab}^\Rsm,\Po_{ab}^\Lsm\,, \Po_{\theta\,ab},\beta_a)\,.
\eeq

We now summarize our study of the restrictions imposed on $n$-point densities by kinematical symmetries of the Poincar\'e superalgebra as follows.

\noindent \ibf) (Anti)commutation relations  between the dynamical generators $P^-$, $Q^{-\Rsm,\Lsm}$, $J^{-\Rsm,\Lsm}$ and the kinematical generators $P^{\Rsm,\Lsm}$,  $P^+$, $Q^{+\Lsm}$ lead to delta-functions in \rf{14052019-man02-08},\rf{14052019-man02-09} and hence imply the conservation laws for the  momenta $p_a^{\Rsm,\Lsm}$, $\beta_a$ and the Grassmann momenta $p_{\theta_a}$.

\noindent \iibf) (Anti)commutation relations  between the dynamical generators $P^-$, $Q^{-\Rsm,\Lsm}$, $J^{-\Rsm,\Lsm}$ and the kinematical generators $J^{+-}$, $J^{\Rsm\Lsm}$ lead to the differential equations given in \rf{14052019-man02-18}-\rf{14052019-man02-24}.

\noindent \iiibf) (Anti)commutation relations between the dynamical generators $P^-$, $Q^{-\Rsm,\Lsm}$, $J^{-\Rsm,\Lsm}$ and the kinematical generators $J^{+\Rsm,\Lsm}$, $Q^{+\Rsm}$ tell us that the $n$-point densities $p_\smpn^-$, $q_\smpn^{-\Rsm,\Lsm}$, $j_\smpn^{-\Rsm,\Lsm}$  turn out to be dependent of the momenta $\Po_{ab}^{\Rsm,\Lsm}$, $\Po_{\theta ab}$ \rf{14052019-man02-25} in place of the respective momenta $p_a^{\Rsm,\Lsm}$, $p_{\theta_a}$.

\noindent \ivbf) Using the conservation laws for the momenta $p_a^\Rsm$, $\beta_a$ it is easy to check  that there are only $n-2$ independent momenta $\Po_{ab}^\Rsm$ \rf{14052019-man02-25}. For example, for $n=3$, there is only one independent $\Po^\Rsm$ (see relations below). The same holds true for the momenta $\Po_{ab}^\Lsm$ and $\Po_{ab\, \theta}$.

\newsection{ \large Kinematical and dynamical restrictions on cubic vertices and light-cone gauge dynamical principle } \label{sec-05}

We now restrict our attention to cubic vertices. First, we represent kinematical $J^{+-}$, $J^{\Rsm\Lsm}$ symmetry equations \rf{14052019-man02-18}-\rf{14052019-man02-24} in terms of the momenta $\Po_{ab}^{\Rsm,\Lsm}$ and $\Po_{ab\, \theta}$. Second, we find restrictions imposed by dynamical symmetries.  Finally, we formulate light-cone gauge dynamical principle and present the complete system equations required to determine the cubic vertices uniquely.

\noindent {\bf Kinematical symmetries of the cubic densities}. Taking into account the momentum conservation laws
\be  \label{17052019-man02-01}
p_1^\Rsm + p_2^\Rsm + p_3^\Rsm = 0\,, \quad p_1^\Lsm + p_2^\Lsm + p_3^\Lsm = 0\,, \quad \beta_1 +\beta_2 +\beta_3 =0 \,,\quad p_{\theta_1} + p_{\theta_2} + p_{\theta_3}=0\,,
\ee
we verify that $\Po_{12}^{\Rsm,\Lsm}$, $\Po_{23}^{\Rsm,\Lsm}$, $\Po_{31}^{\Rsm,\Lsm}$ and Grassmann momenta $\Po_{\theta\, 12}$, $\Po_{\theta\, 23}$, $\Po_{\theta\, 31}$ are expressed in terms of new momenta $\Po^{\Rsm,\Lsm}$, $\Po_\theta$,
\be \label{17052019-man02-02}
\Po_{12}^{\Rsm,\Lsm} =\Po_{23}^{\Rsm,\Lsm} = \Po_{31}^{\Rsm,\Lsm} = \Po^{\Rsm,\Lsm} \,,\qquad
\Po_{\theta\, 12} =\Po_{\theta\, 23} = \Po_{\theta\, 31} = \Po_\theta \,,
\ee
where the new momenta $\Po^{\Rsm,\Lsm}$ and $\Po_\theta$ are defined as
\beq
&& \Po^\Rsm \equiv \frac{1}{3}\sum_{a=1,2,3} \betach_a p_a^\Rsm\,, \qquad \Po^\Lsm \equiv \frac{1}{3} \sum_{a=1,2,3} \betach_a p_a^\Lsm\,, \qquad
\nonumber\\
\label{17052019-man02-04} && \Po_\theta \equiv \frac{1}{3}\sum_{a=1,2,3} \betach_a p_{\theta_a}\,, \qquad
\betach_a\equiv \beta_{a+1}-\beta_{a+2}\,, \quad \beta_a\equiv
\beta_{a+3}\,.
\eeq
We find it convenient to use the momenta \rf{17052019-man02-04} because these momenta are manifestly invariant under cyclic permutations of the external line indices
$1,2,3$. Therefore, using the simplified notation for the densities,
\be  \label{17052019-man02-05-05}
p_\smp3^- = p_{\lambda_1\lambda_2\lambda_3}^- \,, \qquad  q_\smp3^{-\Rsm,\Lsm} = q_{\lambda_1\lambda_2\lambda_3}^{-\Rsm,\Lsm}\,, \qquad  j_\smp3^{-\Rsm,\Lsm}  = j_{\lambda_1\lambda_2\lambda_3}^{-\Rsm,\Lsm}\,,
\ee
we note then the our cubic densities $p_\smp3^-$, $q_\smp3^{-\Rsm,\Lsm}$, and $j_\smp3^{-\Rsm,\Lsm}$ are functions of
the momenta $\beta_a$, $\Po^{\Rsm,\Lsm}$, the Grassmann momentum $\Po_\theta$ and the helicities $\lambda_1$, $\lambda_2$, $\lambda_3$,
\beq
\label{17052019-man02-06} && p_\smp3^- = p_{\lambda_1\lambda_2\lambda_3}^-(\Po^\Rsm,\Po^\Lsm,\Po_\theta, \beta_a)\,, \qquad  q_\smp3^{-\Rsm,\Lsm} = q_{\lambda_1\lambda_2\lambda_3}^{-\Rsm,\Lsm}(\Po^\Rsm,\Po^\Lsm,\Po_\theta, \beta_a)\,, \quad
\nonumber\\
&& j_\smp3^{-\Rsm,\Lsm} = j_{\lambda_1\lambda_2\lambda_3}^{-\Rsm,\Lsm}(\Po^\Rsm,\Po^\Lsm,\Po_\theta, \beta_a)\,.
\eeq
Thus we see that the momenta $p_a^{\Rsm,\Lsm}$ and $p_{\theta_a}$  enter cubic densities \rf{17052019-man02-06} through the respective momenta $\Po^{\Rsm,\Lsm}$ and $\Po_\theta$. This feature of the cubic densities simplifies considerably the study of kinematical symmetry equations presented in \rf{14052019-man02-18}-\rf{14052019-man02-24}. We now represent  equations \rf{14052019-man02-18}-\rf{14052019-man02-24} in terms of cubic densities \rf{17052019-man02-06}.

\noindent {\bf $J^{+-}$-symmetry equations}: Using representation for cubic densities in \rf{17052019-man02-06}, we find that, for $n=3$,  equations \rf{14052019-man02-18},\rf{14052019-man02-19} take the form
\be \label{17052019-man02-07}
(\Jbf^{+-} - 1) p_\smp3^- = 0\,,  \hspace{0.7cm} (\Jbf^{+-} - \frac{3}{2}) q_\smp3^{-\Rsm,\Lsm} = 0\,, \hspace{0.7cm} (\Jbf^{+-} - 1) j_\smp3^{-\Rsm,\Lsm} = 0\,,\qquad
\ee
where $\Jbf^{+-}$ stands for an operator defined as
\beq
\label{17052019-man02-08} && \Jbf^{+-} \equiv    N_{\Po^\Rsm} + N_{\Po^\Lsm}+ \frac{3}{2} N_{\Po_\theta} +  \sum_{a=1,2,3}  \big( \beta_a \partial_{\beta_a} + \half e_{\lambda_a}\big)\,,
\\
\label{17052019-man02-09} && N_{\Po^\Rsm} \equiv \Po^\Rsm\partial_{\Po^\Rsm}\,, \hspace{1cm} N_{\Po^\Lsm} \equiv \Po^\Lsm \partial_{\Po^\Lsm}\,, \hspace{1cm} N_{\Po_\theta} \equiv \Po_\theta \partial_{\Po_\theta}\,.\qquad
\eeq
\noindent {\bf $J^{\Rsm\Lsm}$-symmetry equations}: Using the representation for the cubic densities in \rf{17052019-man02-06}, we find that, for $n=3$,  equations \rf{14052019-man02-20}-\rf{14052019-man02-24} take the form
\beq
\label{17052019-man02-10} &&  (\Jbf^{\Rsm\Lsm} + 1)p_\smp3^- =0\,, \hspace{0.7cm} (\Jbf^{\Rsm\Lsm} + \half) q_\smp3^{-\Rsm} =0\,,\hspace{0.7cm}  (\Jbf^{\Rsm\Lsm} + \frac{3}{2}) q_\smp3^{-\Lsm} =0\,,
\nonumber\\
&& \hspace{4cm} \Jbf^{\Rsm\Lsm} j_\smp3^{-\Rsm} =0\,,   \hspace{1.8cm}  (\Jbf^{\Rsm\Lsm} + 2 ) j_\smp3^{-\Lsm} =0\,,\qquad
\eeq
where $\Jbf^{\Rsm\Lsm}$ stands for an operator defined as
\be \label{17052019-man02-11}
\Jbf^{\Rsm\Lsm} \equiv    N_{\Po^\Rsm} -  N_{\Po^\Lsm} - \half N_{\Po_\theta} +\Mbf_\lambda\,, \qquad \Mbf_\lambda \equiv \sum_{a=1,2,3} \lambda_a\,,
\ee
and we use the notation in \rf{17052019-man02-09}.

We now proceed with studying the restrictions imposed by dynamical symmetries.

\noindent {\bf Dynamical symmetries of the cubic densities}. In this paper, restrictions on the interaction vertices imposed by (anti)commutation relations between the dynamical generators will be referred to as dynamical symmetry restrictions.
We now discuss restrictions imposed on cubic interaction vertices by the dynamical symmetries of the Poincar\'e superalgebra. In other words, we consider the
(anti)commutators
\beq
\label{17052019-man02-12} && [P^-,J^{-\Rsm,\Lsm}]=0\,, \hspace{1.8cm}  [P^-,Q^{-\Rsm,\Lsm}]=0\,,
\\
\label{17052019-man02-13} && [J^{-\Rsm},J^{-\Lsm}]=0\,, \hspace{1.9cm}  [Q^{-\Rsm,\Lsm},J^{-\Lsm}]=0\,,     \hspace{1.2cm}    [Q^{-\Rsm,\Lsm},J^{-\Rsm}]=0\,,
\\
\label{17052019-man02-14} && \{Q^{-\Rsm},Q^{-\Lsm} \} = - P^-\,,\hspace{1cm}
\{Q^{-\Rsm},Q^{-\Rsm} \} = 0\,, \hspace{1cm} \{ Q^{-\Lsm},Q^{-\Lsm} \} = 0\,.
\eeq
Let us first consider the commutation relations given in \rf{17052019-man02-12}. In the cubic approximation, commutation relations given in \rf{17052019-man02-12} can be represented as
\be \label{17052019-man02-15}
[P_\smpt^- ,J_\smp3^{-\Rsm}] + [P_\smp3^-,J_\smpt^{-\Rsm}]=0\,, \qquad [P_\smpt^-,Q_\smp3^{-\Rsm,\Lsm}] + [P_\smp3^-, Q_\smpt^{-\Rsm,\Lsm}]=0\,.
\ee
Using equations \rf{17052019-man02-15}, we find the following representation for the densities $q_\smp3^{-\Rsm,\Lsm}$ and $j_\smp3^{-\Rsm,\Lsm}$ in terms of the cubic vertex $p_\smp3^-$,
\beq
\label{17052019-man02-16} && q_\smp3^{-\Rsm} = - \frac{\epsilon \Po_\theta}{\Po^\Lsm}  p_\smp3^-\,, \hspace{2.2cm}   q_\smp3^{-\Lsm} = \frac{\epsilon \beta}{\Po^\Rsm} \partial_{\Po_\theta} p_\smp3^-\,,
\\
\label{17052019-man02-17} && j_\smp3^{-\Rsm}  = -\frac{\beta}{ \Po^\Rsm \Po^\Lsm } \Jbf^{-\Rsm} p_\smp3^- \,, \hspace{1cm} j_\smp3^{-\Lsm}  = -\frac{\beta}{ \Po^\Rsm \Po^\Lsm } \Jbf^{-\Lsm} p_\smp3^- \,,
\\
\label{17052019-man02-17-a1} && \hspace{1cm} \epsilon  =(-)^{\Ebf_\lambda }\,, \hspace{1cm} \Ebf_\lambda \equiv \sum_{a=1,2,3} e_{\lambda_a}\,,
\eeq
where $\Jbf^{-\Rsm}$, $\Jbf^{-\Lsm}$ stand for operators defined by the relations
\beq
\label{17052019-man02-18} && \Jbf^{-\Rsm} =   \frac{\Po^\Rsm}{\beta}(-\No_\beta + \Mo_\lambda + \half \Eo_{\lambda+\half} )\,,
\\
\label{17052019-man02-19} && \Jbf^{-\Lsm}  =    \frac{\Po^\Lsm}{\beta} (-\No_\beta - \Mo_\lambda + \half \Eo_{\lambda+\half} )\,,
\\
\label{17052019-man02-20} && \hspace{1.3cm} \No_\beta = \frac{1}{3}\sum_{a=1,2,3}\betach_a \beta_a\partial_{\beta_a}\,, \hspace{1cm} \beta \equiv \beta_1\beta_2\beta_3\,,
\\
\label{17052019-man02-21} &&  \hspace{1.3cm} \Mo_\lambda = \frac{1}{3}\sum_{a=1,2,3}\betach_a\lambda_a\,, \hspace{1.5cm} \Eo_{\lambda+\half} = \frac{1}{3}\sum_{a=1,2,3}\betach_a e_{\lambda_a+\half} \,,
\eeq
while $e_\lambda$ appearing in \rf{17052019-man02-17-a1},\rf{17052019-man02-21} is defined in \rf{15052019-man02-20}.

Using expressions for $q_\smp3^{-\Rsm,\Lsm}$ and $j_\smp3^{-\Rsm,\Lsm}$ given in \rf{17052019-man02-16},\rf{17052019-man02-17}, we verify that all the kinematical symmetry equations for $q_\smp3^{-\Rsm,\Lsm}$ and $j_\smp3^{-\Rsm,\Lsm}$ given in \rf{17052019-man02-07},\rf{17052019-man02-10} are satisfied automatically  provided the vertex $p_\smp3^-$ satisfies the kinematical symmetry equations for $p_\smp3^-$ in \rf{17052019-man02-07},\rf{17052019-man02-10}. Using expressions for $q_\smp3^{-\Rsm,\Lsm}$ and $j_\smp3^{-\Rsm,\Lsm}$ given in \rf{17052019-man02-16},\rf{17052019-man02-17}, we also verify that, in the cubic approximation, all (anti)commutation relations given in \rf{17052019-man02-13},\rf{17052019-man02-14} are satisfied automatically.
This is to say that, in the cubic approximation, we checked that the kinematical symmetry equations for $p_\smp3^-$ in \rf{17052019-man02-07},\rf{17052019-man02-10} and equations \rf{17052019-man02-16},\rf{17052019-man02-17} provide the complete list of equations which are obtainable from all the (anti)commutation relations of the Poincar\'e superalgebra.

\noindent {\bf Light-cone gauge dynamical principle}. The kinematical symmetry equations for $p_\smp3^-$ in \rf{17052019-man02-07},\rf{17052019-man02-10} and the equations \rf{17052019-man02-16},\rf{17052019-man02-17}  do not allow to determine the cubic vertex $p_\smp3^-$ unambiguously. In order to determine the cubic vertex $p_\smp3^-$ unambiguously we should impose some additional restrictions on the cubic vertex $p_\smp3^-$. We will refer to such additional restrictions as light-cone gauge dynamical principle. The light-cone gauge dynamical principle is formulated as follows:

\noindent \ibf) The densities $p_\smp3^-$, $q_\smp3^{-\Rsm,\Lsm}$, $j_\smp3^{-\Rsm,\Lsm}$ are required to be polynomial in the momenta  $\Po^\Rsm$, $\Po^\Lsm$;

\noindent \iibf) The cubic vertex $p_\smp3^-$ is required to satisfy the restriction
\be \label{17052019-man02-22}
p_\smp3^-  \ne  \Po^\Rsm\Po^\Lsm W\,, \quad W \ \hbox{is polynomial in } \Po^\Rsm,\Po^\Lsm\,.
\ee
The restriction on the vertex $p_\smp3^-$ in \rf{17052019-man02-22} is
related to the freedom of field redefinitions. We recall that, upon field redefinitions, the vertex $p_\smp3^-$ is changed by terms proportional to $\Po^\Rsm \Po^\Lsm$ (see, e.g., Appendix B in Ref.\cite{Metsaev:2005ar}). This implies that ignoring requirement \rf{17052019-man02-22} leads to cubic vertices which can be removed by field redefinitions. As we are interested in the cubic vertices $p_\smp3^-$
that cannot be removed by field redefinitions, we impose the requirement \rf{17052019-man02-22}. Note also the assumption \ibf) is the light-cone counterpart of locality condition commonly used in gauge invariant and Lorentz covariant formulations.

\noindent {\bf Complete system of equations for cubic vertex}. To summarize the discussion in this section, we note that, for the cubic vertex given by
\be  \label{17052019-man02-23}
p_\smp3^- = p_{\lambda_1\lambda_2\lambda_3}^-(\Po^\Rsm,\Po^\Lsm,\Po_\theta, \beta_a)
\ee
the complete system of equations which remain to be analysed is given by
\beq
\label{17052019-man02-24}  && (\Jbf^{+-} - 1) p_\smp3^- =0 \,, \hspace{4.7cm} \hbox{kinematical } \  J^{+-}-\hbox{ symmetry};
\\
\label{17052019-man02-25}  &&  (\Jbf^{\Rsm\Lsm} + 1)p_\smp3^- = 0\,, \hspace{4.8cm} \hbox{kinematical } \  J^{\Rsm\Lsm}-\hbox{ symmetry};
\\
\label{17052019-man02-26} && j_\smp3^{-\Rsm,\Lsm} = - \frac{\beta}{\Po^\Rsm\Po^\Lsm}\Jbf^{-\Rsm,\Lsm} p_\smp3^- \,, \hspace{3.3cm} \hbox{ dynamical } P^-, J^{-\Rsm,\Lsm} \hbox{ symmetries };\qquad
\\
\label{17052019-man02-27} && q_\smp3^{-\Rsm} = - \frac{\epsilon \Po_\theta}{\Po^\Lsm}  p_\smp3^-\,, \hspace{0.4cm}   q_\smp3^{-\Lsm} = \frac{\epsilon \beta}{\Po^\Rsm} \partial_{\Po_\theta} p_\smp3^-\,, \hspace{1.5cm} \hbox{ dynamical } P^-, Q^{-\Rsm,\Lsm} \hbox{ symmetries };\qquad
\\
&& \hspace{3cm} \hbox{ \it Light-cone gauge dynamical principle:}
\nonumber\\
\label{17052019-man02-28} && p_\smp3^-\,, \ q_\smp3^{-\Rsm,\Lsm}\,,  \ j_\smp3^{-\Rsm,\Lsm} \hspace{0.5cm} \hbox{ are polynomial in } \Po^\Rsm, \Po^\Lsm;
\\
\label{17052019-man02-29} && p_\smp3^- \ne \Po^\Rsm\Po^\Lsm W, \hspace{1cm} W  \hbox{ is polynomial in } \Po^\Rsm, \Po^\Lsm; \qquad
\eeq
Equations given in \rf{17052019-man02-24}-\rf{17052019-man02-29} constitute our basic complete
system of equations which allow us to determine the cubic vertex $p_\smp3^-$ and densities $q_\smp3^{-\Rsm,\Lsm}$, $j_\smp3^{-\Rsm,\Lsm}$ uniquely. Differential operators $\Jbf^{+-}$, $\Jbf^{\Rsm\Lsm}$, $\Jbf^{-\Rsm,\Lsm}$ and quantity $\epsilon$ entering our basic equations are given in  \rf{17052019-man02-08},\rf{17052019-man02-11}, \rf{17052019-man02-18},\rf{17052019-man02-19} and \rf{17052019-man02-17-a1} respectively.
Considering the super Yang-Mills and supergravity theories, we can verify that our basic equations in \rf{17052019-man02-24}-\rf{17052019-man02-29} allow us to determine the cubic interaction vertices of those theories unambiguously (up to coupling constants). It seems then reasonable to use our equations for studying the cubic vertices of arbitrary spin supersymmetric theories.

\newsection{ \large Cubic interaction vertices } \label{sec-06}

We now present the solution to our basic equations for densities given in \rf{17052019-man02-24}-\rf{17052019-man02-29}.

General solution for the cubic interaction vertex $p_{\lambda_1\lambda_2\lambda_3}^-$, the supercharge  densities $q_{\lambda_1\lambda_2\lambda_3}^{-\Rsm,\Lsm}$, and the angular momentum densities $j_{\lambda_1\lambda_2\lambda_3}^{-\Rsm,\Lsm}$ we found is given by (for some details of the derivation, see Appendix C)
\beq
\label{19052019-man02-01} && p_{\lambda_1\lambda_2\lambda_3}^- = V_{\lambda_1\lambda_2\lambda_3} + \Vb_{\lambda_1\lambda_2\lambda_3}\,,
\\
\label{19052019-man02-02} && \hspace{1.5cm} V_{\lambda_1\lambda_2\lambda_3} =  C^{\lambda_1\lambda_2\lambda_3} (\Po^\Lsm)^{\Mbf_\lambda + 1} \prod_{a=1,2,3} \beta_a^{-\lambda_a  -  \half e_{\lambda_a} }\,,
\\
\label{19052019-man02-03} && \hspace{1.5cm} \Vb_{\lambda_1\lambda_2\lambda_3}  = \Cb^{\lambda_1\lambda_2\lambda_3} (\Po^\Rsm)^{- \Mbf_\lambda  - \half}\, \Po_\theta  \prod_{a=1,2,3} \beta_a^{\lambda_a  -  \half e_{\lambda_a}  }\,,
\\
\label{19052019-man02-04} && q_{\lambda_1\lambda_2\lambda_3}^{-\Rsm} = - C^{\lambda_1\lambda_2\lambda_3} (\Po^\Lsm)^{ \Mbf_\lambda }\, \Po_\theta \prod_{a=1,2,3} \beta_a^{-\lambda_a  -  \half e_{\lambda_a} }\,,
\\
\label{19052019-man02-05} && q_{\lambda_1\lambda_2\lambda_3}^{-\Lsm} =  - \Cb^{\lambda_1\lambda_2\lambda_3} (\Po^\Rsm)^{- \Mbf_\lambda-\frac{3}{2}} \prod_{a=1,2,3} \beta_a^{\lambda_a + 1 -  \half e_{\lambda_a} }\,,
\\
\label{19052019-man02-06} && j_{\lambda_1\lambda_2\lambda_3}^{-\Rsm} =  - 2 C^{\lambda_1\lambda_2\lambda_3} \Mo_\lambda (\Po^\Lsm)^{ \Mbf_\lambda} \prod_{a=1,2,3} \beta_a^{-\lambda_a  -  \half e_{\lambda_a} }\,,
\\
\label{19052019-man02-07} && j_{\lambda_1\lambda_2\lambda_3}^{-\Lsm} =    2\Cb^{\lambda_1\lambda_2\lambda_3} \Mo_\lambda (\Po^\Rsm)^{- \Mbf_\lambda-\frac{3}{2} }\, \Po_\theta \prod_{a=1,2,3} \beta_a^{\lambda_a - \half e_{\lambda_a} }\,,
\eeq
where, in \rf{19052019-man02-01}-\rf{19052019-man02-07} and below, we use the notation
\be \label{19052019-man02-08}
\Mbf_\lambda = \sum_{a=1,2,3}\lambda_a\,, \hspace{1cm} \Mo_\lambda = \frac{1}{3}\sum_{a=1,2,3}\betach_a \lambda_a\,, \qquad \Ebf_\lambda = \sum_{a=1,2,3} e_{\lambda_a}\,,
\ee
while the symbol $e_\lambda$ and the momenta $\Po^{\Rsm,\Lsm}$, $\Po_\theta$, $\betach_a$ are defined in \rf{15052019-man02-20} and \rf{17052019-man02-04} respectively.
Quantities $C^{\lambda_1\lambda_2\lambda_3}$, $\Cb^{\lambda_1\lambda_2\lambda_3}$ appearing in \rf{19052019-man02-01}-\rf{19052019-man02-07} stand for coupling constants which, in general, depend on the helicities $\lambda_1$, $\lambda_2$, $\lambda_3$. These coupling constants are nontrivial for the following values of the helicities:
\beq
\label{19052019-man02-09}  && C^{\lambda_1,\lambda_2,\lambda_3} \ne 0\,, \hspace{ 1cm } \hbox{ for  $\Mbf_\lambda \geq   0$ \hspace{0.7cm} and \ \ $\Mbf_\lambda$ - integer } \,,
\\
\label{19052019-man02-10} && \Cb^{\lambda_1,\lambda_2,\lambda_3 } \ne 0 \,, \hspace{ 1cm } \hbox{ for  $\Mbf_\lambda \leq -\frac{3}{2} $ \hspace{0.3cm} and \ \ $\Mbf_\lambda$ - half-integer } \,,
\\
\label{19052019-man02-11} && C^{\lambda_1,\lambda_2,\lambda_3 *} =  (-)^{\Mbf_\lambda + e_{\lambda_2} + 1} \Cb^{-\lambda_1-\half,-\lambda_2-\half,-\lambda_3-\half}\,.
\eeq
Let us discuss restrictions in \rf{19052019-man02-09}-\rf{19052019-man02-11} in turn.

\noindent \ibf) Restrictions on $C^{\lambda_1,\lambda_2,\lambda_3}$ and $\Mbf_\lambda$ in \rf{19052019-man02-09} are obtained by requiring the densities \rf{19052019-man02-02},\rf{19052019-man02-04},\rf{19052019-man02-06} to be polynomial in  $\Po^\Lsm$, while restrictions on $\Cb^{\lambda_1,\lambda_2,\lambda_3}$ and $\Mbf_\lambda$ in \rf{19052019-man02-10} are obtained by requiring the densities \rf{19052019-man02-03},\rf{19052019-man02-05},\rf{19052019-man02-07} to be polynomial in  $\Po^\Rsm$.

\noindent \iibf) Restrictions for $\Mbf_\lambda$ to be integer in \rf{19052019-man02-09} and half-integer in \rf{19052019-man02-10} can also obtained by requiring the Hamiltonian $P_\smp3^-$ to be Grassmann even.
Namely, taking into account the Grassmann parities of the vertices $V_{\lambda_1,\lambda_2,\lambda_3}$, $\Vb_{\lambda_1,\lambda_2,\lambda_3}$, the integration measure $d\Gamma_\smp3^{p_\theta}$, and product of the three superfields $\Theta_{\lambda_1\lambda_2\lambda_3}^*$ \rf{14052019-man02-15},
\be \label{19052019-man02-12}
\GP(V_{\lambda_1,\lambda_2,\lambda_3}) = 0,\quad \GP(\Vb_{\lambda_1,\lambda_2,\lambda_3}) =1\,, \quad
\GP(d\Gamma_\smp3^{p_\theta}) = 0,\quad \GP(\Theta_{\lambda_1\lambda_2\lambda_3}^*) = \Ebf_\lambda \,,
\ee
and requiring $\GP(P_\smp3^-)=0$, we get the restrictions
\be \label{19052019-man02-13}
\GP(d\Gamma_\smp3^{p_\theta} \Theta_{\lambda_1\lambda_2\lambda_3}^* V^{\lambda_1,\lambda_2,\lambda_3}) = 0,\quad   \GP(d\Gamma_\smp3^{p_\theta} \Theta_{\lambda_1\lambda_2\lambda_3}^* \Vb^{\lambda_1,\lambda_2,\lambda_3}) =0\,,
\ee
which amount to restrictions for $\Mbf_\lambda$ to be integer in \rf{19052019-man02-09} and half-integer in \rf{19052019-man02-10}.

\noindent \iiibf) Requiring the cubic Hamiltonian $P_\smp3^-$ to be hermitian and using relation \rf{04062019-man02-16} in Appendix B, we get the restrictions for coupling constants given in \rf{19052019-man02-11}.

Expressions \rf{19052019-man02-01}-\rf{19052019-man02-03} provide the momentum superspace representation for all cubic vertices, while relations in \rf{19052019-man02-09}-\rf{19052019-man02-11} provide the classification of such vertices.

\noindent {\bf Incorporation of internal symmetry}. Let the algebra $o(\Nsf)$ be an internal symmetry algebra. We incorporate an internal symmetry into our model in the following way.

\noindent \ibf) In place fields $\phi_\lambda$, $\psi_\lambda$ \rf{15052019-man02-04}-\rf{15052019-man02-06}, we introduce  fields $\phi_\lambda^{\asf\bsf}$, $\psi_\lambda^{\asf\bsf}$, while,
in place of the superfields $\Theta_\lambda$, $\Theta_\lambda^*$ \rf{16052019-man02-07},\rf{16052019-man02-22}, we introduce superfields $\Theta_\lambda^{\asf\bsf}$, $\Theta_\lambda^{*\asf\bsf}$, where indices $\asf,\bsf$ are  matrix indices of the $o(\Nsf)$ algebra, $\asf,\bsf=1,\ldots,\Nsf$. By definition, new superfields satisfy the algebraic constraints
\be \label{29052019-man02-05}
\Theta_\lambda^{\asf\bsf} = (-)^{\lambda-\half e_\lambda} \Theta_\lambda^{\bsf\asf} \,, \qquad \Theta_\lambda^{*\asf\bsf} = (-)^{\lambda+\half e_\lambda} \Theta_\lambda^{*\bsf\asf} \,,
\ee
where $e_\lambda$ is given in \rf{15052019-man02-20}. Using the relation $(-)^{2\lambda \pm e_\lambda} = 1$, we verify that constraints \rf{29052019-man02-05} are consistent. The superfields $\Theta_\lambda^{\asf\bsf}$ and $\Theta_\lambda^{*\asf\bsf}$ are not independent of each other. By analogy with \rf{16052019-man02-23}, we have the relation
\be
\Theta_{-\lambda}^{*\asf\bsf}(-p,-p_\theta) =  (-)^{e_\lambda} \Theta_\lambda^{\asf\bsf}(p,p_\theta)\,.
\ee
From \rf{29052019-man02-05}, we learn that the $\Theta_\lambda^{\asf\bsf}$ is symmetric in $\asf\bsf$ for even $s$ in \rf{16052019-man02-07}, while, for odd $s$ in \rf{16052019-man02-07}, the $\Theta_\lambda^{\asf\bsf}$ is antisymmetric in $\asf\bsf$. Also, from \rf{29052019-man02-05}, we learn that the $\Theta_\lambda^{*\asf\bsf}$ is antisymmetric in $\asf\bsf$ for even $s$ in \rf{16052019-man02-22}, while, for odd $s$ in \rf{16052019-man02-22}, the $\Theta_\lambda^{*\asf\bsf}$ is symmetric in $\asf\bsf$.
Note that relations \rf{29052019-man02-05} imply that $\phi_\lambda^{\asf\bsf}$, $\psi_\lambda^{\asf\bsf}$ are symmetric in $\asf\bsf$ for even $s$ in \rf{15052019-man02-04}-\rf{15052019-man02-06},
while for odd $s$ in \rf{15052019-man02-04},\rf{15052019-man02-05} the $\phi_\lambda^{\asf\bsf}$, $\psi_\lambda^{\asf\bsf}$ are antisymmetric in $\asf\bsf$.
The hermicity conditions \rf{15052019-man02-11} take the form $(\phi_\lambda^{\asf\bsf}(p))^\dagger = \phi_{-\lambda}^{\asf\bsf}(-p)$, $(\psi_\lambda^{\asf\bsf}(p))^\dagger = \psi_{-\lambda}^{\asf\bsf}(-p)$.

\noindent \iibf) In \rf{16052019-man02-24}, the expressions $\Theta_{\lambda-(1/2)}^* \Theta_\lambda$ are replaced by  $\Theta_{\lambda-(1/2)}^{*\asf\bsf} \Theta_\lambda^{\asf\bsf}$,
while, in the cubic vertices, the expressions $\Theta_{\lambda_1}^*
\Theta_{\lambda_2}^* \Theta_{\lambda_3}^*$ are replaced by the trace
$\Theta_{\lambda_1}^{*\asf\bsf}
\Theta_{\lambda_2}^{*\bsf\csf} \Theta_{\lambda_3}^{*\csf\asf}$.

\noindent \iiibf) In place of graded commutator \rf{16052019-man02-26}, we use
\beq
\label{29052019-man02-06} &&  [\Theta_\lambda^{\asf\bsf}(p,p_\theta),\Theta_{\lambda'}^{*\asf'\bsf'}(p',p_\theta')]_\pm = \frac{(-)^{ e_{\lambda+\half} } }{2\beta}  \Pi_\lambda^{\asf\bsf,\asf'\bsf'} \delta^3(p-p')\delta(p_\theta-p_\theta')  \delta_{\lambda-\lambda',\half}\,,
\\
\label{29052019-man02-07} && \Pi_\lambda^{\asf\bsf,\asf'\bsf'} \equiv \half\big( \delta^{\asf\asf'} \delta^{\bsf\bsf'} + (-)^{\lambda -\half e_\lambda}  \delta^{\asf\bsf'} \delta^{\bsf\asf'} \big)\,, \qquad \Pi_\lambda^{\asf\bsf,\asf'\bsf'} \Pi_\lambda^{\asf'\bsf',\csf\esf} = \Pi_\lambda^{\asf\bsf,\csf\esf}\,,
\eeq
where the second relation in \rf{29052019-man02-07} is verified by using the relation $(-)^{2\lambda - e_\lambda} = 1$.

\noindent {\bf Cubic Hamiltonian in terms of component fields}. To make our results more transparent and pragmatic we now discuss an explicit representation for the cubic Hamiltonian $P_\smp3^-$ \rf{14052019-man02-02} in terms of component fields \rf{15052019-man02-04}-\rf{15052019-man02-06} and demonstrate explicitly powers of momenta in our cubic vertices. To this end we restrict our attention to interaction of three superfields $\Theta_{\lambda_1}^*$, $\Theta_{\lambda_2}^*$, $\Theta_{\lambda_3}^*$ and represent the corresponding cubic Hamiltonian $P_\smp3^-$ \rf{14052019-man02-02} with $p_{\lambda_1\lambda_2\lambda_3}^-$ \rf{19052019-man02-01} as follows
\beq
\label{19052019-man02-17} && P_\smp3^-(\Theta_{\lambda_1},\Theta_{\lambda_2}\Theta_{\lambda_3}) = \int d\Gamma_\smp3^p\,\, C^{\lambda_1\lambda_2\lambda_3} \Vbf^{ \Theta_{\lambda_1},\Theta_{\lambda_2}\Theta_{\lambda_3} } + h.c.
\\
\label{19052019-man02-18} && C^{\lambda_1\lambda_2\lambda_3} \Vbf^{\Theta_{\lambda_1}\Theta_{\lambda_2}\Theta_{\lambda_3} } \equiv  \int  d\Gamma_\smp3^{p_\theta}\,\, \Theta_{\lambda_1\lambda_2\lambda_3}^* V_{\lambda_1\lambda_2\lambda_3}\,,
\eeq
where the integration measures $d\Gamma_\smp3^p$, $d\Gamma_\smp3^{p_\theta}$ are obtained by setting $n=3$ in \rf{14052019-man02-08},\rf{14052019-man02-09}. It is the vertex $\Vbf^{\Theta_{\lambda_1}\Theta_{\lambda_2}\Theta_{\lambda_3} }$ \rf{19052019-man02-18} that we refer to as the vertex in terms of the component fields.

From \rf{19052019-man02-02},\rf{19052019-man02-18}, we see that the vertex $\Vbf^{\Theta_{\lambda_1}\Theta_{\lambda_2}\Theta_{\lambda_3} }$ is nontrivial if and only if the $\Theta_{\lambda_1\lambda_2\lambda_3}^*$-term is Grassmann even. Classification of all such $\Theta_{\lambda_1\lambda_2\lambda_3}^*$-terms may be found in Appendix B (see relations \rf{20052019-man02-01}-\rf{20052019-man02-04}).
Here we present those Grassmann even $\Theta_{\lambda_1\lambda_2\lambda_3}^*$-terms given  in \rf{20052019-man02-01}-\rf{20052019-man02-04} that respect restriction on $\Mbf_\lambda$ \rf{19052019-man02-09}. We classify all relevant Grassmann even $\Theta_{\lambda_1\lambda_2\lambda_3}^*$-terms as follows:
\beq
\label{19052019-man02-20} && \hspace{-2.4cm} {\small\bf Cases \ 1ab}: \hspace{1cm} \Phi_{s_1-\half}^*\Phi_{s_2-\half}^*\Psi_{s_3}^*\,, \hspace{1cm} \Psi_{s_1}^*\Psi_{s_2}^*\Psi_{s_3}^*;
\\
\label{19052019-man02-21} && \hspace{-2.4cm} {\small\bf Cases \ 2abc}: \hspace{0.7cm} \Phi_{s_1-\half}^*\Phi_{s_2-\half}^*\Phi_{-s_3}^*\,, \hspace{0.9cm} \Phi_{s_1-\half}^*\Psi_{s_2}^*\Psi_{-s_3-\half}^*\,, \hspace{0.7cm} \Psi_{s_1}^*\Psi_{s_2}^*\Phi_{-s_3}^*;
\\
\label{19052019-man02-22} && \hspace{-2.4cm} {\small \bf Cases \ 3abc}: \hspace{0.7cm} \Phi_{s_1-\half}^*\Phi_{-s_2}^*\Psi_{-s_3-\half}^*\,, \hspace{0.7cm} \Psi_{s_1}^*\Phi_{-s_2}^*\Phi_{-s_3}^*\,, \hspace{1.4cm} \Psi_{s_1}^*\Psi_{-s_2-\half}^*\Psi_{-s_3-\half}^*\,.
\eeq
Now, all that remains is to plug expressions for $\Phi_{s-\half}^*$, $\Phi_{-s}^*$, and $\Psi_s^*$, $\Psi_{-s-\half}^*$ \rf{16052019-man02-18}-\rf{16052019-man02-20} into \rf{19052019-man02-18} and make integration over the Grassmann momenta $p_{\theta_1}$, $p_{\theta_2}$, $p_{\theta_3}$.
In order to simplify our presentation of the vertices $\Vbf^{\Theta_{\lambda_1}\Theta_{\lambda_2}\Theta_{\lambda_3} }$, we collect the fields $\phi_\lambda^\dagger$ and $\psi_\lambda^\dagger$ into a field $\theta_\lambda^\dagger = \phi_\lambda^\dagger, \psi_\lambda^\dagger$ and use the following shortcut for a product of the component fields
\be \label{19052019-man02-19}
V^{\theta_{\lambda_1} \theta_{\lambda_2}\theta_{\lambda_3} } \equiv \frac{\theta_{\lambda_1}^\dagger(p_1) \theta_{\lambda_2}^\dagger(p_2) \theta_{\lambda_3}^\dagger (p_3) }{\beta_1^{\lambda_1+ \half e_{\lambda_1} }\beta_2^{\lambda_2 + \half e_{\lambda_2} }\beta_3^{\lambda_3 + \half e_{\lambda_3} } } \hspace{1.6cm} \hbox{ for } \lambda_1+\lambda_2+\lambda_3 > 0\,. \qquad
\ee
Now, using notation \rf{19052019-man02-19}, we present expressions for the vertices $\Vbf^{\Theta_{\lambda_1}\Theta_{\lambda_2}\Theta_{\lambda_3} }$ \rf{19052019-man02-18} corresponding to the cases \rf{19052019-man02-20}-\rf{19052019-man02-22} in turn.
In due course we show explicitly powers of momentum $\Po^\Lsm$ appearing in the vertices.

\noindent {\bf Cases 1ab}. Vertices of powers $(\Po^\Lsm)^{s_1+s_2+s_3}$ and $(\Po^\Lsm)^{s_1+s_2+s_3+1}$ :
{\small
\beq
&& \hspace{-2cm} \Vbf^{\Phi_{s_1-\half}\Phi_{s_2-\half}\Psi_{s_3}} =
\left( - V^{\phi_{s_1}\phi_{s_2}\psi_{s_3}} + V^{ \phi_{s_1} \phi_{s_2-\half} \psi_{s_3+\half} } +  V^{\phi_{s_1-\half}\phi_{s_2} \psi_{s_3+\half} }\right) (\Po^\Lsm)^{s_1+s_2+s_3}\,,
\nonumber\\
\label{19052019-man02-23}  && \hspace{4cm} s_1\geq 1\,, \quad s_2\geq 1 \,, \quad s_3\geq 0\,, \hspace{2.7cm} ({\bf 1a})
\\
&& \hspace{-2cm} \Vbf^{ \Psi_{s_1}\Psi_{s_2}\Psi_{s_3} } = \left(
V^{\psi_{s_1}\psi_{s_2+\half}\psi_{s_3+\half} } - V^{\psi_{s_1+\half}\psi_{s_2}\psi_{s_3+\half} } +  V^{\psi_{s_1+\half}\psi_{s_2+\half}\psi_{s_3} }\right) (\Po^\Lsm)^{s_1+s_2+s_3+1}\,,
\nonumber\\
\label{19052019-man02-24} && \hspace{4cm} s_1\geq 0\,, \quad s_2\geq 0 \,, \quad s_3\geq 0\,, \hspace{2.7cm} ({\bf 1b})
\eeq
}

\noindent {\bf Cases 2abc}. Vertices of powers $(\Po^\Lsm)^{s_1+s_2-s_3}$ and $(\Po^\Lsm)^{s_1+s_2-s_3+1}$ :
{\small
\beq
&& \hspace{-1.7cm} \Vbf^{\Phi_{s_1-\half}\Phi_{s_2-\half}\Phi_{-s_3} } = \left(
- V^{\phi_{s_1}\phi_{s_2}\phi_{-s_3} } +  V^{\phi_{s_1}\phi_{s_2-\half} \phi_{-s_3+\half} } + V^{\phi_{s_1-\half}\phi_{s_2}\phi_{-s_3+\half} }\right) (\Po^\Lsm)^{s_1+s_2-s_3}\,,
\nonumber\\
\label{19052019-man02-26} && \hspace{2.5cm} s_1\geq 1\,, \quad s_2\geq 1 \,, \quad s_3\geq 1\,, \qquad s_1+s_2-s_3\geq 1\,, \hspace{1.7cm} ({\bf 2a})
\\
&& \hspace{-1.7cm} \Vbf^{ \Phi_{s_1-\half}\Psi_{s_2}\Psi_{-s_3-\half} } = \left(
V^{\phi_{s_1}\psi_{s_2}\psi_{-s_3} } +  V^{\phi_{s_1} \psi_{s_2+\half} \psi_{-s_3-\half} } - V^{\phi_{s_1-\half}\psi_{s_2+\half}\psi_{-s_3} }\right) (\Po^\Lsm)^{s_1+s_2-s_3}\,,
\nonumber\\
\label{19052019-man02-27} && \hspace{2.5cm} s_1\geq 1\,, \quad s_2\geq 0 \,, \quad s_3\geq 0\,, \qquad s_1+s_2-s_3\geq 1\,, \hspace{1.7cm} ({\bf 2b})
\\
&& \hspace{-1.7cm} \Vbf^{\Psi_{s_1}\Psi_{s_2}\Phi_{-s_3} }  = \left(
V^{\psi_{s_1+\half}\psi_{s_2+\half}\phi_{-s_3} } +  V^{ \psi_{s_1} \psi_{s_2+\half} \phi_{-s_3+\half} } - V^{ \psi_{s_1+\half} \psi_{s_2} \phi_{-s_3+\half}} \right) (\Po^\Lsm)^{s_1+s_2-s_3+1}\,,
\nonumber\\
\label{19052019-man02-28} && \hspace{2.5cm} s_1\geq 0\,, \quad s_2\geq 0 \,, \quad s_3\geq 1\,, \qquad s_1+s_2-s_3\geq 0\,. \hspace{1.7cm} ({\bf 2c})
\eeq
}

\noindent {\bf Cases 3abc}.  Vertices of powers $(\Po^\Lsm)^{s_1-s_2-s_3}$ and $(\Po^\Lsm)^{s_1-s_2-s_3+1}$ :
{\small
$$
{}\quad\Vbf^{\Phi_{s_1-\half}\Phi_{-s_2}\Psi_{-s_3-\half} } = \big(
V^{\phi_{s_1}\phi_{-s_2}\psi{-s_3} } +  V^{ \phi_{s_1}\phi_{-s_2+\half}\psi_{-s_3-\half} } - V^{\phi_{s_1-\half},\phi_{-s_2+\half}\psi_{-s_3} }\big) (\Po^\Lsm)^{s_1-s_2-s_3}\,, \hspace{1cm}
$$
\be \label{19052019-man02-29}
\hspace{3.7cm} s_1\geq 1\,, \quad s_2\geq 1 \,, \quad s_3\geq 0\,, \qquad s_1-s_2-s_3\geq 1\,, \hspace{1.7cm} ({\bf 3a})
\ee

\vspace{-0.5cm}
$$
{}\quad \Vbf^{\Psi_{s_1}\Phi_{-s_2}\Phi_{-s_3} } = \big(
V^{\psi_{s_1}\phi_{-s_2+\half}\phi_{-s_3+\half} } -  V^{\psi_{s_1+\half}\phi_{-s_2}\phi_{-s_3+\half} } + V^{\psi_{s_1+\half}\phi_{-s_2+\half}\phi_{-s_3} }\big) (\Po^\Lsm)^{s_1-s_2-s_3+1}\,, \hspace{1cm}
$$
\be \label{19052019-man02-30}
\hspace{3.7cm} s_1\geq 0\,, \quad s_2\geq 1 \,, \quad s_3\geq 1\,, \qquad s_1 - s_2-s_3\geq 0\,, \hspace{1.7cm} ({\bf 3b})
\ee

\vspace{-0.5cm}
$$
\quad \Vbf^{\Psi_{s_1}\Psi_{-s_2-\half}\Psi_{-s_3-\half} } = -\big(
 V^{\psi_{s_1}\psi_{-s_2}\psi_{-s_3} } + V^{\psi_{s_1+\half}\psi_{-s_2-\half}\psi_{-s_3} } + V^{\psi_{s_1+\half}\psi_{-s_2}\psi_{-s_3-\half} }\big) (\Po^\Lsm)^{s_1-s_2-s_3}\,, \hspace{1cm}
$$
\be \label{19052019-man02-31}
\hspace{3.7cm} s_1\geq 0\,, \quad s_2\geq 0 \,, \quad s_3\geq 0\,, \qquad s_1 - s_2-s_3\geq 1\,. \hspace{1.7cm} ({\bf 3c})
\ee
}

To summarize, our superspace cubic vertices given in \rf{19052019-man02-01} with restrictions
\rf{19052019-man02-09}-\rf{19052019-man02-11} provide the full list of cubic interaction vertices that can be constructed for integer and half-integer spin massless $N=1$ supermultiplets. Representation of our vertices \rf{19052019-man02-01} in terms of the component fields is given in \rf{19052019-man02-23}-\rf{19052019-man02-31}.%
\footnote{ In the framework of Lorentz covariant approach, the recent extensive study of cubic vertices of higher-spin $N=1$ supermultiplets by using gauge invariant supercurrents, may be found in Refs.\cite{Buchbinder:2017nuc}-\cite{Gates:2019cnl}. Lorentz covariant superfield formulations of free $N=1$ supermultiplets in $4d$ flat space were studied in Refs.\cite{Kuzenko:1993jq}.}

In \rf{19052019-man02-23}-\rf{19052019-man02-31}, we classified our vertices focusing on the powers of the momentum $\Po^\Lsm$ appearing in the vertices. Focusing on the number of integer spin supermultiplets $(s,s-\half)$ (superfield $\Phi$) and half-integer spin supermultiplets $(s+\half,s)$  (superfield $\Psi$) appearing in the vertices, we can reclassify our vertices.  Namely, focusing on the number of superfields $\Phi$, $\Psi$ we can  symbolically represent our classification \rf{19052019-man02-23}-\rf{19052019-man02-31} as follows.
\beq
&&\hspace{-2cm} \hbox{ Three integer spin supermultiplets}: \hspace{3.5cm} \Phi\Phi\Phi\hbox{-- vertices in }  ({\bf 2a})
\\[-3pt]
&& \hspace{-2cm} \hbox{ Two integer and one half-integer spin supermultiplets}: \hspace{0.2cm}  \Phi\Phi\Psi\hbox{--vertices in } ({\bf 1a}) ({\bf 3a}) ({\bf 3b})
\\[-3pt]
&& \hspace{-2cm} \hbox{ One integer and two half-integer spin supermultiplets}:  \hspace{0.3cm} \Phi\Psi\Psi\hbox{--vertices in }  ({\bf 2b}) ({\bf 2c})
\\[-3pt]
&& \hspace{-2cm} \hbox{ Three half-integer spin supermultiplets}:  \hspace{2.7cm} \Psi\Psi\Psi\hbox{--vertices in } ({\bf 1b})({\bf 3c})
\eeq

To illustrate our result let us consider particular cases from the list in \rf{19052019-man02-23}-\rf{19052019-man02-31}.

\noindent \ibf) For the particular case $s_1=0$, $s_2=0$, $s_3=0$, the vertex \rf{19052019-man02-24} takes the form
\be \label{19052019-man02-25}
\Vbf^{\Psi_0\Psi_0\Psi_0} =
\left( V^{\psi_0\psi_\half\psi_\half} - V^{ \psi_\half \psi_0 \psi_\half } +  V^{\psi_\half \psi_\half \psi_0 }\right) \Po^\Lsm\,,  \hspace{1cm} \hbox{ WZ susy model}.
\ee
Vertex \rf{19052019-man02-25} provides light-cone description of the well known   supersymmetric WZ model.

\noindent \iibf) For the particular cases $s_1=1$, $s_2=1$, $s_3=1$ and
$s_1=2$, $s_2=2$, $s_3=2$, vertex \rf{19052019-man02-26} provides  light-cone gauge description of cubic vertices of the respective super Yang-Mills and supergravity theories,
\beq
\label{29052019-man02-01} && \hspace{-1.7cm} \Vbf^{\Phi_{\half}\Phi_{\half}\Phi_{-1} } = \left(
- V^{\phi_1\phi_1\phi_{-1} } +  V^{\phi_1\phi_{\half} \phi_{- \half} } + V^{\phi_{ \half}\phi_{s_2}\phi_{- \half} }\right) \Po^\Lsm \,, \hspace{1cm} \hbox{ super YM theory};
\\
\label{29052019-man02-02} && \hspace{-1.7cm} \Vbf^{\Phi_{\frac{3}{2}}\Phi_{\frac{3}{2}}\Phi_{-2} } = \left(
- V^{\phi_2\phi_2\phi_{-2} } +  V^{\phi_2 \phi_{ \frac{3}{2}   } \phi_{- \frac{3}{2}   } } + V^{\phi_{\frac{3}{2}}\phi_2\phi_{-\frac{3}{2}} }\right) (\Po^\Lsm)^2\,, \hspace{0.7cm} \hbox{ supergravity};
\eeq

\noindent \iiibf) If, for the particular cases $s_1=1$, $s_2=1$, $s_3=1$ and $s_1=2$, $s_2=2$, $s_3=2$, we consider vertices that involve {\it only} integer spin supermultiplets \rf{19052019-man02-26}, then we do not find vertices with three momenta for spin-1 fields and vertices with six momenta for spin-2 fields. This reflects the well known fact that, by using {\it only integer spin} supermultiplets $(1,\half)$ and $(2,\frac{3}{2})$, it is not possible to build supersymmetric extensions of $F_{YM}^3$ terms and $R^3$ terms, where $F_{YM}$ and $R$ stand for the respective field strength of YM field and Riemann tensor. The same happens for cubic vertices that involves {\it only} the half-integer spin supermultiplets \rf{19052019-man02-24},\rf{19052019-man02-31}.
It turns out however, that, if, for the particular cases $s_1=1$, $s_2=1$, $s_3=1$ and $s_1=2$, $s_2=2$, $s_3=2$, we consider vertices that involves {\it both} the integer and half integer spin supermultiplets \rf{19052019-man02-23}, then we find vertices with three momenta for spin-1 fields and vertices with six momenta for spin-2 fields given by
\beq
\label{29052019-man02-03} && \hspace{-2cm} \Vbf^{\Phi_{\half}\Phi_{\half}\Psi_1} =
\left( - V^{\phi_1 \phi_1 \psi_1 } + V^{ \phi_1 \phi_\half \psi_{ \frac{3}{2}  } } +  V^{\phi_\half \phi_1 \psi_{\frac{3}{2}} }\right) (\Po^\Lsm)^3\,, \hspace{0.4cm} \hbox{ $F_{YM}^3$ super YM-like theory};
\\
\label{29052019-man02-04} && \hspace{-2cm} \Vbf^{\Phi_{\frac{3}{2}}\Phi_{\frac{3}{2}}\Psi_2} =
\left( - V^{\phi_2\phi_2\psi_2} + V^{ \phi_2 \phi_{\frac{3}{2}} \psi_{\frac{5}{2}} } +  V^{\phi_{\frac{3}{2}}\phi_2 \psi_{\frac{5}{2}} }\right) (\Po^\Lsm)^6\,, \hspace{0.4cm} \hbox{ $R^3$ supergravity-like theory};
\eeq
Thus, at least in the cubic approximation, by using two integer supermultiplets $(1,\half)$ and one half integer spin supermultiplet $(\frac{3}{2},1)$, we can build supersymmetric extension of $F_{YM}^3$-terms \rf{29052019-man02-03}, while, by using two integer supermultiplets $(2,\frac{3}{2})$ and one half integer spin supermultiplet $(\frac{5}{2},2)$, we can build
supersymmetric extension of $R^3$-terms \rf{29052019-man02-04}.

\noindent {\bf Interrelations between number of derivatives in light-cone gauge and covariant approaches}.  To make our results more useful and helpful for those readers who prefer Lorentz covariant formulations  we now discuss a correspondence between number of momenta (transverse derivatives) appearing in our light-cone gauge cubic vertices and number of momenta (derivatives) appearing in the corresponding  Lorentz covariant theory. Using shortcuts $B$ and $F$  for the respective massless bosonic and massless fermionic fields, we write schematically a cubic Lagrangian of Lorentz covariant theory $\LL_\cov$ and related light-cone gauge cubic Lagrangian $\LL_\lc$ as follows
\beq
\label{19052019-man02-14} && \LL_\cov  =   P^{K_{BBB}^\cov} B B  B   + P^{K_{FFB}^\cov} FF B\,,
\\
\label{19052019-man02-15} && \LL_\lc  =   \Po^{K_{BBB}^\lc} B B  B   + \Po^{K_{FFB}^\lc} FF B\,.
\eeq
In \rf{19052019-man02-14}, $P$ stands for momenta (derivatives), while $K_{BBB}^\cov$ and $K_{FFB}^\cov$ denote numbers of momenta $P$ (derivatives) entering cubic vertices in metric-like Lorentz covariant formulation. Accordingly in \rf{19052019-man02-15}, the $\Po$ stands for the momenta $\Po^\Rsm,\Po^\Lsm$ (transverse derivatives), while
$K_{BBB}^\lc$ and $K_{FFB}^\lc$ denote numbers of momenta  $\Po$ (transverse derivatives) entering our light-cone gauge cubic vertices.
We note the following relations for the numbers of the momenta (derivatives)%
\footnote{Relations \rf{19052019-man02-16} are valid for metric-like Lorentz covariant formulations that do not involve auxiliary fields. In general, for Lorentz covariant formulations that involve auxiliary fields, relations \rf{19052019-man02-16} might break down.}
\be \label{19052019-man02-16}
K_{BBB}^\cov = K_{BBB}^\lc\,, \hspace{1cm}  K_{FFB}^\cov = K_{FFB}^\lc -1\,.
\ee
Now using \rf{19052019-man02-16} and our classification for the light-cone gauge vertices, we propose a classification of covariant vertices. We classify covariant vertices focusing on number of integer spin supermultiplets $(s,s-\half)$ (superfield $\Phi$) and half-integer spin supermultiplets $(s+\half,s)$  (superfield $\Psi$) appearing in the vertices. This is to say that we represent the classification of light-cone gauge vertices  \rf{19052019-man02-23}-\rf{19052019-man02-31}  in terms of the corresponding covariant cubic vertices as follows:
\beq
&& \hbox{ \bf Three integer spin supermultiplets} \ (\Phi\Phi\Phi-\hbox{ vertices}):
\nonumber\\
&& (s_1,s_1-\half)\hbox{-}(s_2,s_2-\half)\hbox{-}(s_3,s_3-\half),   \quad  s_1\geq 1\,, \ s_2\geq 1\,, \ s_3 \geq 1\,,
\nonumber\\
\label{29052019-man02-19} {\bf (2a)} && K_{BBB}^\cov =  s_1 + s_2 - s_3\,, \hspace{0.4cm} K_{FFB}^\cov =  s_1 + s_2 -  s_3 - 1\,,  \hspace{0.3cm} s_1 + s_2 - s_3 \geq  1\,;  \qquad
\\
&& \hbox{ \bf Two integer and one half-integer spin supermultiplets}  \ (\Phi\Phi\Psi-\hbox{ vertices}):
\nonumber\\
&& (s_1,s_1-\half)\hbox{-}(s_2,s_2-\half)\hbox{-}(s_3+\half,s_3),   \quad  s_1\geq 1\,, \ s_2\geq 1\,, \ s_3 \geq 0\,,
\nonumber\\
\label{29052019-man02-20}  {\bf (1a)} && K_{BBB}^\cov =  s_1 + s_2 + s_3\,, \hspace{0.4cm} K_{FFB}^\cov =  s_1 + s_2 +  s_3 - 1\,,  \hspace{0.3cm} s_1 + s_2 + s_3 \geq 1\,,  \qquad
\\
\label{29052019-man02-21} {\bf (3a)} && K_{BBB}^\cov =  s_1 - s_2 -  s_3 \,,  \hspace{0.4cm}    K_{FFB}^\cov =  s_1 - s_2 -  s_3 - 1\,,  \hspace{0.3cm} s_1 - s_2 - s_3 \geq 1\,;  \qquad
\\
&& (s_1+\half,s_1)\hbox{-}(s_2,s_2-\half)\hbox{-}(s_3,s_3-\half),   \quad  s_1\geq 0\,, \ s_2\geq 1\,, \ s_3 \geq 1;
\nonumber\\
\label{29052019-man02-22} {\bf (3b)} && K_{FFB}^\cov =  s_1 - s_2 -  s_3 \,,  \hspace{0.3cm} s_1 - s_2 - s_3 \geq 0\,,  \qquad
\\
&& \hbox{\bf One integer spin and two half integer supermultiplets}  \ (\Phi \Psi\Psi- \hbox{ vertices}):
\nonumber\\
&& (s_1,s_1-\half)\hbox{-}(s_2+\half,s_2)\hbox{-}(s_3+\half,s_3),   \quad  s_1\geq 1\,, \ s_2\geq 0\,, \ s_3 \geq 0\,,
\nonumber\\
\label{29052019-man02-23} {\bf (2b)} && K_{BBB}^\cov =  s_1 + s_2 -  s_3 \,,  \hspace{0.3cm} K_{FFB}^\cov =  s_1 + s_2 -  s_3 - 1\,,  \hspace{0.3cm} s_1 + s_2 - s_3 \geq 1\,,  \qquad
\\
&& (s_1+\half,s_1)\hbox{-}(s_2+\half,s_2)\hbox{-}(s_3,s_3-\half),   \quad  s_1\geq 0\,, \ s_2\geq 0\,, \ s_3 \geq 1;
\nonumber\\
\label{29052019-man02-24} {\bf (2c)} &&  K_{FFB}^\cov =  s_1 + s_2 -  s_3 \,,  \hspace{0.3cm} s_1 + s_2 - s_3 \geq 0\,;  \qquad
\\
&& \hbox{ \bf Three half-integer spin supermultiplets}  \ (\Psi\Psi\Psi-\hbox{ vertices}):
\nonumber\\
&& (s_1+\half,s_1)\hbox{-}(s_2+\half,s_2)\hbox{-}(s_3+\half,s_3),   \quad  s_1\geq 0\,, \ s_2\geq 0\,, \ s_3 \geq 0\,,
\nonumber\\
\label{29052019-man02-25} {\bf (1b)} &&  K_{FFB}^\cov =  s_1 + s_2 + s_3 \,, \hspace{1cm} \hspace{0.3cm} s_1 + s_2 + s_3 \geq 0\,,  \qquad
\\
\label{29052019-man02-26} {\bf (3c)} && K_{BBB}^\cov =  s_1 - s_2 - s_3\,, \hspace{0.4cm} K_{FFB}^\cov =  s_1 - s_2 -  s_3 - 1\,,  \hspace{0.3cm} s_1 - s_2 - s_3 \geq 1\,.  \qquad
\eeq
In the left column in \rf{29052019-man02-19}-\rf{29052019-man02-26}, we use the labels to show explicitly the correspondence between the classification for covariant vertices in \rf{29052019-man02-19}-\rf{29052019-man02-26} and the one for the light-cone gauge vertices in \rf{19052019-man02-23}-\rf{19052019-man02-31}.

\noindent {\bf On problem of manifestly Lorentz covariant formulation of light-cone gauge vertices}. Vertices given in \rf{29052019-man02-19}-\rf{29052019-man02-21},\rf{29052019-man02-23},\rf{29052019-man02-26} provide the supersymmetric extension for all cubic vertices for massless bosonic fields in the $4d$ flat space presented in Ref.\cite{Bengtsson:1986kh}. Note however that a manifestly Lorentz covariant formulation of some light-cone gauge vertices in Ref.\cite{Bengtsson:1986kh} is not available so far. Therefore a manifestly Lorentz covariant formulation of some our supersymmetric light-cone gauge vertices is not easy problem. For the reader convenience, we now discuss those vertices in \rf{29052019-man02-19}-\rf{29052019-man02-21},\rf{29052019-man02-23},\rf{29052019-man02-26} that, as we expect, can be converted into manifestly Lorentz covariant form in relatively straightforward way. To this end, let us consider cubic vertex for massless bosonic spin  $s_1$-, $s_2$-, $s_3$- fields having $k$ powers of derivatives. Results in Refs.\cite{Metsaev:2005ar,Conde:2016izb} imply that, in $4d$ space, if
$k$ take the values
\be \label{23072019-man02-01}
k = \sbf,\, \sbf-2 s_\minrm, \qquad \sbf \equiv s_1 + s_2 + s_3\,, \qquad s_\minrm \equiv \min(s_1,s_2,s_3)\,,
\ee
then the cubic vertex can be represented into manifestly Lorentz covariant form.%
\footnote{ In Ref.\cite{Metsaev:2005ar}, we noted that parity-even cubic vertices for light-cone gauge massless fields in $R^{d-1,1}$, $d$-arbitrary, lead to two parity-even vertices with $k$ as in \rf{23072019-man02-01} when $d=4$. In Ref.\cite{Conde:2016izb}, it was observed that Lorentz covariant parity-odd cubic vertices for on-shell massless field in $4d$ have also $k$ as in \rf{23072019-man02-01}. Manifestly Lorentz covariant description of all parity-even cubic vertices for off-shell massless fields in dimensions $d\geq4$ is given in Refs.\cite{Fotopoulos:2010ay,Metsaev:2012uy}.}
Now, making assumption that bosonic vertices with $k$ as in \rf{23072019-man02-01} allow manifestly Lorentz covariant supersymmetric extension, we can fix vertices that can straightforwardly be converted into manifestly Lorentz covariant form by considering the equation
\be \label{23072019-man02-02}
K_{BBB}^\cov = k\,,
\ee
where values of $k$ are given in \rf{23072019-man02-01}, while values of $K_{BBB}^\cov$ are given in \rf{29052019-man02-19}-\rf{29052019-man02-21},\rf{29052019-man02-23},\rf{29052019-man02-26}.
For some vertices, equation \rf{23072019-man02-02} imposes additional restrictions on allowed values of $s_1$, $s_2$, $s_3$. If a vertex satisfies equation \rf{23072019-man02-02}, then such vertex can be represented into manifestly Lorentz covariant form, while, if  a vertex does not satisfy equation \rf{23072019-man02-02}, then manifestly Lorentz covariant formulation of such vertex is not easy problem. We now present result of analysis of solutions of equation \rf{23072019-man02-02} for the vertices \rf{29052019-man02-19}-\rf{29052019-man02-21},\rf{29052019-man02-23},\rf{29052019-man02-26}.

\noindent \ibf) Vertices \rf{29052019-man02-19},\rf{29052019-man02-23} with the additional restrictions $s_3\leq s_1$, $s_3\leq s_2$ can be represented in manifestly Lorentz covariant form, while manifestly Lorentz covariant formulation of all remaining vertices in \rf{29052019-man02-19},\rf{29052019-man02-23} is not easy problem.

\noindent \iibf) All vertices in \rf{29052019-man02-20} can be translated into manifestly Lorentz covariant form.

\noindent \iiibf) Manifestly Lorentz covariant formulation of all vertices in \rf{29052019-man02-21} is not easy problem.

\noindent \ivbf) Vertices \rf{29052019-man02-26} with the additional restrictions $s_2=0$, $s_3=0$ can be converted into manifestly Lorentz covariant form, while manifestly Lorentz covariant formulation of all remaining vertices in \rf{29052019-man02-26} is not easy problem.

Vertices \rf{29052019-man02-22},\rf{29052019-man02-24},\rf{29052019-man02-25} describe arbitrary spin WZ-like supersymmetric models. We expect that manifestly Lorentz covariant formulation of all vertices in \rf{29052019-man02-22} is not easy problem, while all vertices in \rf{29052019-man02-25} can be translated into manifestly Lorentz covariant form. It is likely that vertices \rf{29052019-man02-24} with additional restrictions $s_3\leq s_1$, $s_3\leq s_2$ can be represented into manifestly Lorentz covariant form, while manifestly Lorentz covariant formulation of all remaining vertices in \rf{29052019-man02-24} is not easy problem.

\noindent {\bf Motivation for study of both the integer and half-integer spin supermultiplets in flat space}. We can try to restrict our attention to the study of supersymmetric higher-spin theory that involves {\it only integer spin supermultiplets}. It turns out that {\it such theory does not exist in the flat space}.
Our  arguments are as follows. Consider vertex \rf{19052019-man02-26} with $s_1=s$, $s_2=s$, $s_3=s$, $s> 2$. In Ref.\cite{Metsaev:1991mt}, for the case of bosonic theories, we demonstrated that in order to respect some restrictions on such vertex which appear at the quartic order one needs to use, among other things, the vertices of powers $(\Po^\Lsm)^{s_1+s_2+s_3}$. From the expressions in \rf{19052019-man02-23}, we see however that, in  our supersymmetric theory, such vertices can be build if and only if we use both the integer and half-integer spin supermultiplets.
Thus, the $N=1$ supersymmetry in higher-spin theory in the flat $4d$ space requires the use of both the integer and half-integer spin supermultiplets. In other words, we should use the chain of fields that involves each helicity twice. Appearance of such chain of fields in the supersymmetric higher-spin theory in $AdS_4$ space is the well known fact \cite{Konstein:1989ij} (see also Ref.\cite{Engquist:2002vr}).

Finally, we conjecture that the solution for coupling constants $C^{\lambda_1\lambda_2\lambda_3}$ in Refs.\cite{Metsaev:1991mt,Metsaev:1991nb}  can be generalized to the case of $N=1$ supersymmetric higher-spin theory considered in this paper as follows
\be  \label{19052019-man02-32}
C^{\lambda_1\lambda_2\lambda_3} = \frac{ g (-)^{ e_{\lambda_2} } k^{\lambda_1+\lambda_2+\lambda_3} }{ (\lambda_1 + \lambda_2 + \lambda_3)! }\,,
\ee
where $e_\lambda$ is defined in \rf{15052019-man02-20}. In \rf{19052019-man02-32}, the $g$ is a dimensionless coupling constant, while $k$ is some dimensionfull complex-valued parameter in general. The $g$ and $k$ do not depend on the helicities. For the supersymmetric theory with hermitian Hamiltonian, the constants $\Cb^{\lambda_1\lambda_2\lambda_3}$ are fixed by the relation in \rf{19052019-man02-11}, while, for supersymmetric generalization of the chiral higher-spin theory in Ref.\cite{Ponomarev:2016lrm}, we should set $\Cb^{\lambda_1\lambda_2\lambda_3}=0$. For the bosonic truncation of our supersymmetric model, solution \rf{19052019-man02-32} amounts to the one in Refs.\cite{Metsaev:1991mt,Metsaev:1991nb}.
Solution \rf{19052019-man02-32} can be used for discussion at least the following two supersymmetric higher-spin field models in the flat $4d$ space.

\noindent \ibf) Field content of the first model is given in \rf{15052019-man02-04}-\rf{15052019-man02-06} and described by the superfields $\Theta_\lambda$ with all values $\lambda$ in \rf{16052019-man02-04}. For this model, the superfields $\Theta_\lambda$ \rf{16052019-man02-04} are matrices of the internal symmetry $o(\Nsf)$ algebra denoted as $\Theta_\lambda^{ab}$. By definition, the $\Theta_\lambda^{ab}$ are subject to the algebraic constraint in \rf{29052019-man02-05}.

\noindent \iibf) In the second model, the superfields $\Theta_\lambda$ are singlets of the $o(\Nsf)$ algebra and we use the set of superfields given by
\be \label{30052019-man02-01}
\sum_{ \lambda -\half e_\lambda \in 2 \Zo   } \oplus \,\, \Theta_\lambda\,,
\ee
where the summation is performed over those values of $\lambda$ \rf{16052019-man02-04} that satisfy the restriction $(-)^{\lambda - \half e_\lambda } = 1$. In terms of the superfields $\Phi_\lambda$, $\Psi_\lambda$, \rf{16052019-man02-07}, the set of superfields in \rf{30052019-man02-01} can be presented as
\be \label{30052019-man02-02}
\sum_{ n = 1   }^\infty    \oplus \,\,  \Phi_{2n} \oplus \Phi_{-2n+\half} +  \sum_{n=0}^\infty \oplus \,\, \Psi_{2n+\half} \oplus \Psi_{-2n}\,.
\ee
In terms of the component fields, using notation $(s,s-\half)$ and $(s+\half,s)$ for the respective supermultiplets in \rf{15052019-man02-04} and \rf{15052019-man02-05},\rf{15052019-man02-06}, we represent the field content of the second model as
\be \label{30052019-man02-03}
\sum_{ n = 1   }^\infty    \oplus \,\,  (2n,2n-\half) +  \sum_{n=0}^\infty \oplus \,\, (2n+\half,2n)\,.
\ee
Appearance of such two models in the $N=1$ supersymmetric higher-spin theory in AdS space is the well known fact. We note that, in terms of the superfields $\Theta_\lambda^*$, $\Phi_\lambda^*$, $\Psi_\lambda^*$, relations \rf{30052019-man02-01},\rf{30052019-man02-02} take the form
\beq
&& \sum_{  \lambda + \half e_\lambda \in 2 \Zo   }  \oplus \,\, \Theta_\lambda^*\,, %
\\
&& \sum_{ n = 1   }^\infty    \oplus \,\,  \Phi_{2n-\half}^* \oplus \Phi_{-2n}^* +  \sum_{n=0}^\infty \oplus \,\, \Psi_{2n}^* \oplus \Psi_{-2n-\half}^*\,.
\eeq

\newsection{ \large Conclusions}\label{concl}

In this paper, we used light-cone gauge formalism for studying the $N=1$ integer spin and half-integer supermultiplets in the flat $4d$ space. For such supermultiplets, we developed the light-cone gauge formulation in terms of the unconstrained superfields. We used our superfield formulation to build the full list of the cubic vertices that describe interactions of massless integer and half-integer spin supermultiplets. Taking into account powers of momenta appearing in our cubic interaction vertices, we concluded that the integer spin supermultiplets alone are not enough for the studying the full theory of massless $N=1$ interacting supermultiplets in the flat $4d$ space.
For the studying the full $N=1$ supersymmetric theory of higher-spin massless fields in the flat $4d$ space one needs to use both the integer and half-integer spin supermultiplets. In other words, as compared to bosonic theory of massless higher-spin fields, in supersymmetric theory of higher-spin fields, one needs to use the double set of fields (each helicity occurs twice). In this respect, the supersymmetric theory  of higher-spin massless fields in the $4d$ flat space and the one in $AdS_4$ space are similar.  We believe that results in this paper might be helpful for the  following generalizations and applications.

\medskip
\noindent \ibf) In this paper, we studied supersymmetric {\it massless} higher-spin theory in the flat $4d$ space. Generalization of our results to the case of supersymmetric {\it massive} fields in the flat $4d$ space could be of interest.
We note that all parity invariant cubic vertices for massless and massive arbitrary spin fields in the flat space $R^{d-1,1}$, $d$-arbitrary, were built in Refs.\cite{Metsaev:2005ar,Metsaev:2007rn,Metsaev:2012uy}. Namely, in Refs.\cite{Metsaev:2005ar,Metsaev:2007rn}, we built all parity invariant cubic vertices for massless and massive bosonic and fermionic fields in the framework of light-cone gauge formalism, while, in Ref.\cite{Metsaev:2012uy}, we built all parity invariant cubic vertices for massless and massive bosonic fields in the framework of BRST-BV approach.%
\footnote{ In the framework of various Lorentz covariant approaches, cubic vertices for  massless bosonic fields were studied in Refs.\cite{Fotopoulos:2010ay}. Study of Fermi-Bose couplings in the framework of BV approach may be found in Refs.\cite{Henneaux:2012wg}. Fermi-Bose couplings of fields in $R^{3,1}$ by using the light-cone gauge helicity basis were studied in Ref.\cite{Akshay:2015kxa}. Interesting formulation of fermionic fields is developed in Ref.\cite{Najafizadeh:2018cpu}.  Discussion of various aspects of interacting fields in the framework of BRST approach may be found in Refs.\cite{Bekaert:2005jf,Kaparulin:2019quz}. In the framework of Lorentz covariant approach, parity-odd cubic vertices for higher-spin massless fields in $R^{3,1}$ are considered in Ref.\cite{Conde:2016izb}.}
We expect that light-cone gauge cubic vertices in Refs.\cite{Metsaev:2005ar,Metsaev:2007rn} will be helpful for the studying supersymmetric theories of massless and massive fields.
Discussion of supermultiplets in various dimensions  may be found, e.g., in Ref.\cite{Sorokin:2018djm}. Study of $N=1$ higher-spin massless supermultiplets  via BRST approach may be found in Ref.\cite{Buchbinder:2015kca}, while the $N=1$
massive supermultiplets are investigated in Ref.\cite{Zinoviev:2007js}.

\noindent \iibf) In this paper, we restricted our attention to supersymmetric massless higher-spin theory in {\it the flat $4d$ space}. Gauge invariant formulation of the higher-spin theory in AdS space is well known \cite{Vasiliev:1990en}. Various aspects of supersymmetric higher-spin gauge field theory in AdS space have extensively been studied in the past (see, e.g., Refs.\cite{Konstein:1989ij,Engquist:2002vr,Alkalaev:2002rq}). Generalization of our results to the case of light-cone gauge supersymmetric massless higher-spin fields in {\it $AdS_4$ space} could of great interest. Light-cone gauge cubic interaction vertices of higher-spin massless fields in $AdS_4$ space have recently been obtained in Ref.\cite{Metsaev:2018xip}. We believe therefore that result in this paper and the one in Ref.\cite{Metsaev:2018xip} provide a good starting point for the studying light-cone gauge supersymmetric massless fields in $AdS_4$ space.%
\footnote{ We mention also methods in Refs.\cite{Joung:2011ww} which might be useful for analysis of supersymmetric higher-spin theories in AdS. Study of supersymmetric higher-spin models by using world line methods (see, e.g., Refs.\cite{Alkalaev:1999hi}) could also be of some interest.}

\noindent \iiibf) In the recent time, there has been increasing interest in the studying various  higher-spin theories in three-dimensional flat and AdS spaces (see, e.g., Refs.\cite{Nilsson:2015pua}-\cite{Kessel:2018ugi}  and references therein). We think that  the light-cone gauge approach will simplify considerably the whole analysis of higher-spin massive and conformal fields in three dimensions.%
\footnote{ We recall that higher-spin massless fields do not propagate in three dimensions and these fields are trivial in the light-cone gauge. Therefore, for higher-spin massless fields in three dimensions, the usefulness of the light-cone formulation is questionable.}
The light-cone gauge formulation of massive fields in the flat space is well known (see, e.g., Ref.\cite{Siegel:1988yz}), while the light-cone gauge formulation of massive fields in $AdS_3$ space was developed in Refs.\cite{Metsaev:1999ui,Metsaev:2000qb}. The ordinary-derivative light-cone gauge formulation of free conformal fields was developed in Ref.\cite{Metsaev:2016rpa}. We expect that use of the light-cone formulation in Refs.\cite{Metsaev:2000qb,Metsaev:2016rpa} might be helpful for better understanding of various theories in three dimensions.

\noindent \ivbf) As discussed in Ref.\cite{Skvortsov:2018jea}, the chiral higher-spin model \cite{Ponomarev:2016lrm} is free of one-loop divergencies. Also, the general arguments were given for cancellation of all loop divergencies. Loop diagrams in the chiral higher-spin theory are subset of the ones in the full (non-chiral) higher-spin theory. Therefore, the result in Ref.\cite{Skvortsov:2018jea} is a good sign for the quantum finiteness of full (non-chiral) higher-spin theory. The study of quantum properties of full (non-chiral) higher-spin in flat space theory may be found in Ref.\cite{Ponomarev:2016jqk}. We believe that the light-cone gauge superfield formulation of interacting $N=1$ higher-spin supermultiplets suggested in this paper will bring new interesting novelty in the studying quantum properties of higher-spin theory.

\medskip

{\bf Acknowledgments}. This work was supported by the RFBR Grant No.17-02-00546.

\setcounter{section}{0}\setcounter{subsection}{0}
\appendix{ \large Notation and conventions  }

Grassmann momentum is denoted by $p_\theta$, while the left derivative w.r.t the $p_\theta$ is denoted by $\partial_{p_\theta}$. The integral over the Grassmann momentum $p_\theta$ is defined to be $\int dp_\theta p_\theta =1$. Hermitian conjugation rules for the Grassmann momentum $p_\theta$, the derivative $\partial_{p_\theta}$, and integration measure $dp_\theta$ are assumed to be as follows
\be
p_\theta^\dagger = p_\theta\,, \qquad \partial_{p_\theta}^\dagger  = \partial_{p_\theta}\,,\qquad dp_\theta^\dagger = - dp_\theta\,.
\ee
Ghost parities of the $p_\theta$, $\partial_{p_\theta}$, and measure $dp_\theta$ are given by
\be
\GP({p_\theta) = 1\,, \qquad \GP(\partial_{p_\theta}) = 1\,, \qquad
\GP(dp_\theta}) = 1\,.
\ee

For product of two quantities $A$, $B$ having arbitrary ghost numbers, the hermitian conjugation is defined according to the rule $(AB)^\dagger = B^\dagger A^\dagger$. Various relations for the Berezin integrals are summarized as
\beq
&& \int dp_\theta\, (\partial_{p_\theta} A) B =  (-)^{\epsilon_A+1} \int dp_\theta  A  \partial_{p_\theta} B\,,
\\
&& \int dp_\theta\, ( p_\theta \partial_{p_\theta} A) B =  \int dp_{\theta}  A (1 -  p_\theta \partial_{p_\theta}) B\,, \hspace{1cm}  \epsilon_A \equiv \GP(A)\,, \qquad
\epsilon_B \equiv \GP(B)\,, \qquad
\eeq
where $AB = (-)^{\epsilon_A\epsilon_B} BA$. For $p_{\theta_a}$, $\partial_{p_{\theta_a}}$, and $dp_{\theta_a}$, $a=1,\ldots, n$, entering $n$-point vertices, we assume the conventions
\beq
&& \{ p_{\theta_a}, \partial_{p_{\theta_b}} \} = \delta_{ab} \qquad \int  dp_{\theta_a}  {p_{\theta_b}}  = \delta_{ab}\,,
\\
&& \{p_{\theta_a}, p_{\theta_b} \} = 0 \,, \qquad\{p_{\theta_a}, dp_{\theta_b} \} = 0 \,, \qquad\{\partial_{p_{\theta_a}}, dp_{\theta_b} \} = 0 \,, \qquad \{dp_{\theta_a}, dp_{\theta_b} \} = 0 \,.\qquad
\eeq
Grassmann Dirac delta-function is fixed by the relations
\be
\delta(p_\theta) = p_\theta\,, \qquad \int dp_\theta' \delta(p_\theta'-p_\theta) f(p_\theta') = f(p_\theta)\,.
\ee
Grassmann Fourier transform and its inverse are defined by the relations
\be
F(p_\theta) = \int dp_\theta'e^{\frac{p_\theta' p_\theta}{\beta} } f(p_\theta')\,, \qquad f(p_\theta) = \beta \int dp_\theta'e^{\frac{p_\theta' p_\theta}{\beta} } F(p_\theta')\,.
\ee
Using $ d\Gamma_\smp3^{p_\theta}$ \rf{14052019-man02-09}, we note the various helpful Berezin integrals for 3-point vertices
\beq
\label{04062019-man02-12} && \hspace{-1.5cm} \int d\Gamma_\smp3^{p_\theta}\,\, p_{\theta_a} p_{\theta_{a+1}} = -1\,, \qquad  \int d\Gamma_\smp3^{p_\theta}\,\, p_{\theta_a} \Po_\theta = \beta_a\,, \qquad a=1,2,3\,,
\\
\label{04062019-man02-14} && \hspace{-1.5cm}  \int d\Gamma_\smp3^{p_\theta'}  \exp\big(\sum_{a=1,2,3} \frac{ p_{\theta_a} p_{\theta_a}' }{ \beta_a} \big) = \frac{1}{\beta} \Pbf_\theta \Po_\theta\,,     \hspace{1cm}\int d\Gamma_\smp3^{p_\theta'}\,  \Po_\theta' \exp\big(\sum_{a=1,2,3} \frac{ p_{\theta_a}  p_{\theta_a}' }{ \beta_a} \big) = \Pbf_\theta\,,\qquad
\eeq
where $\Pbf_\theta=p_{\theta_1}+p_{\theta_2}+p_{\theta_3}$ and $\beta =\beta_1\beta_2\beta_3$.

\appendix{ Some properties of superfields   }

To build interaction vertices we find it convenient to use superfields $\Theta_\lambda^*$ defined in \rf{16052019-man02-18}-\rf{16052019-man02-21}.

\noindent {\bf Realizations of Poincar\'e superalgebra on superfield $\Theta_\lambda^*$ in terms of differential operators}:
\beq
\label{04062019-man02-01} && P^\Rsm = - p^\Rsm\,,  \qquad P^\Lsm = - p^\Lsm\,,   \hspace{1.4cm}    P^+ = - \beta\,,\qquad
P^- = -p^-\,, \quad p^- \equiv - \frac{p^\Rsm p^\Lsm}{\beta}\,,\qquad
\\
\label{04062019-man02-02} && J^{+\Rsm}= \irm x^+ P^\Rsm + \partial_{p^\Lsm}\beta\,, \hspace{2.3cm} J^{+\Lsm}= \irm x^+ P^\Lsm + \partial_{p^\Rsm}\beta\,, \
\\
\label{04062019-man02-03} && J^{+-} = \irm x^+P^- + \partial_\beta \beta + M_{-\lambda}^{+-}\,, \hspace{1cm} J^{\Rsm\Lsm} =  p^\Rsm\partial_{p^\Rsm} - p^\Lsm\partial_{p^\Lsm} + M_{-\lambda}^{\Rsm\Lsm}\,,
\\
\label{04062019-man02-04} && J^{-\Rsm} = -\partial_\beta p^\Rsm + \partial_{p^\Lsm} p^-
+ M_{-\lambda}^{\Rsm\Lsm}\frac{p^\Rsm}{\beta} - M_{-\lambda}^{+-} \frac{p^\Rsm}{\beta}\,,
\\
\label{04062019-man02-06} && J^{-\Lsm} = -\partial_\beta p^\Lsm + \partial_{p^\Rsm} p^-
- M_{-\lambda}^{\Rsm\Lsm}\frac{p^\Lsm}{\beta} - M_{-\lambda}^{+-} \frac{p^\Lsm}{\beta}\,,
\\
\label{04062019-man02-07} && \hspace{1.2cm} M_\lambda^{+-} =  \half p_\theta\partial_{p_\theta} - \half e_\lambda\,, \hspace{2cm} M_\lambda^{\Rsm\Lsm}  =   \lambda -\half p_\theta\partial_{p_\theta}\,,
\\
\label{04062019-man02-08} && Q^{+\Rsm} = (-)^{e_\lambda} \beta \partial_{p_\theta}\,, \hspace{2.8cm}  Q^{+\Lsm} = (-)^{ e_{\lambda+\half} } p_\theta\,,
\\
\label{04062019-man02-09} && Q^{-\Rsm} =   (-)^{ e_{\lambda+\half} } \frac{1}{\beta} p^\Rsm p_\theta\,, \hspace{2cm}  Q^{-\Lsm} = (-)^{e_\lambda} p^\Lsm \partial_{p_\theta}\,,
\eeq
where the symbol $e_\lambda$ is defined in \rf{15052019-man02-20}.
Explicit realization of the Poincar\'e superalgebra on superfields $\Phi^*$ and $\Psi^*$ \rf{16052019-man02-18}-\rf{16052019-man02-20} is given in the Table II.

Using relations given in \rf{16052019-man02-24},\rf{16052019-man02-26}, we verify the standard equal-time (anti)commutation relation between the superfields $\Theta_\lambda^*$ and the generators
\be \label{04062019-man02-10}
[\Theta_\lambda^*,G_\smpt]_{\pm} =  G_{\diff,\,\lambda} \Theta_\lambda^* \,,
\ee
where $G_{\diff,\,\lambda}$ are given in \rf{04062019-man02-01}-\rf{04062019-man02-09}. Note that, by using \rf{16052019-man02-23}, the (anti)commutator \rf{16052019-man02-26} can entirely be represented in terms of the superfields $\Theta_\lambda^*$ as
\beq
&& [\Theta_\lambda^*(p,p_\theta),\Theta_{\lambda'}^*(p',p_\theta')]_\pm = - \frac{1}{2\beta}\delta^3(p+p')\delta(p_\theta+p_\theta')\delta_{\lambda+\lambda',-\half}\,,
\eeq

\medskip
\noindent{\small\sf Table II. Realization of supercharges \rf{04062019-man02-08},\rf{04062019-man02-09} on superfields \rf{16052019-man02-18}-\rf{16052019-man02-20}. Realization of the Poincar\'e algebra on superfields \rf{16052019-man02-18}-\rf{16052019-man02-20} is given by relations \rf{04062019-man02-01}-\rf{04062019-man02-06}, where the operators $M_{-\lambda}^{+-}$, $M_{-\lambda}^{\Rsm\Lsm}$ should be replaced by operators $M^{+-}$, $M^{\Rsm\Lsm}$ given in this Table.}
{\small
\begin{center}
\begin{tabular}{|c|c|c|c|c|}
\hline
& & & &
\\[-3mm]
& $\Phi_{s-\half}^*$ & $\Phi_{-s}^*$  & $\Psi_s^*$ & $\Psi_{-s-\half}^*$
\\[2mm]
\hline
&&&&
\\[-3mm]
$Q^{+\Rsm}$ & $ -\beta\partial_{p_\theta}$ & $\beta\partial_{p_\theta}$  & $\beta\partial_{p_\theta}$ & $-\beta\partial_{p_\theta}$
\\[2mm]
\hline
&&&&
\\[-3mm]
$Q^{+\Lsm}$ & $p_\theta$ & $-p_\theta$  & $-p_\theta$ & $p_\theta$
\\[2mm]
\hline
&&&&
\\[-3mm]
$Q^{-\Rsm}$ & $\frac{1}{\beta}p^\Rsm p_\theta$ & $-\frac{1}{\beta}p^\Rsm p_\theta$  & $-\frac{1}{\beta}p^\Rsm p_\theta$ & $\frac{1}{\beta}p^\Rsm p_\theta$
\\[2mm]
\hline
&&&&
\\[-3mm]
$Q^{-\Lsm}$ & $-p^\Lsm\partial_{p_\theta}$ & $ p^\Lsm\partial_{p_\theta}$  & $ p^\Lsm\partial_{p_\theta}$ & $ -p^\Lsm\partial_{p_\theta}$
\\[2mm]
\hline
&&&&
\\[-3mm]
$M^{+-}$ & $ \half p_\theta\partial_{p_\theta}-\half$ & $\half p_\theta\partial_{p_\theta}$ & $\half p_\theta\partial_{p_\theta}$ & $\half p_\theta\partial_{p_\theta}-\half$
\\[2mm]
\hline
&&&&
\\[-3mm]
$M^{\Rsm\Lsm}$ & $ -s-\half p_\theta\partial_{p_\theta}+\half $ & $ s-\half p_\theta\partial_{p_\theta} $  & $-s-\half p_\theta\partial_{p_\theta}$ & $ s-\half p_\theta\partial_{p_\theta}+\half $
\\[2mm]
\hline
\end{tabular}
\end{center}
}

\noindent {\bf Hermitian conjugate of superfields and vertices}. Hermitian conjugate of the superfield $\Theta_\lambda^*$ denoted by $\Theta_\lambda^{*\dagger}$ can be presented as
\be \label{04062019-man02-11}
(\Theta_\lambda^*(p,p_\theta))^\dagger = \beta^{e_\lambda} \int dp_\theta' e^{\frac{p_\theta'p_\theta}{\beta}} \Theta_{-\lambda-\half}^*(-p,p_\theta')\,.\quad
\ee
To prove the restriction for the coupling constants \rf{19052019-man02-11}
we introduce the vertices
\be  \label{04062019-man02-15}
{}\hspace{-0.5cm} v^{\lambda_1\lambda_2\lambda_3} =  (\Po^\Lsm)^{\Mbf_\lambda + 1} \prod_{a=1,2,3} \beta_a^{-\lambda_a  -  \half e_{\lambda_a} }\,,
\hspace{1cm} \vb^{\lambda_1\lambda_2\lambda_3} = (\Po^\Rsm)^{- \Mbf_\lambda  - \half}\, \Po_\theta  \prod_{a=1,2,3} \beta_a^{\lambda_a  -  \half e_{\lambda_a} }\,,
\ee
where $\Mbf_\lambda$ is defined in  \rf{19052019-man02-08}. Using \rf{04062019-man02-14},\rf{04062019-man02-11} considering integer values of $\Mbf_\lambda$, we get the relation
\beq
\label{04062019-man02-16} && \Big(\int d\Gamma_\smp3  \Theta_{\lambda_1\lambda_2\lambda_3}^*  v^{\lambda_1\lambda_2\lambda_3}\Big)^\dagger
\nonumber\\
&& \hspace{1cm} =\,\, (-)^{\Mbf_\lambda + e_{\lambda_2} + 1} \int d\Gamma_\smp3 \Theta_{-\lambda_1-\half,-\lambda_3-\half,-\lambda_3-\half}^* \vb^{-\lambda_1-\half,-\lambda_2-\half,-\lambda_3-\half}\,.\qquad
\eeq
Using \rf{04062019-man02-16}, we see that, requiring the cubic Hamiltonian $P_\smp3^-$ to be hermitian, we get the restrictions for coupling constants given in \rf{19052019-man02-11}.

\noindent {\bf Grassmann even and odd $\Theta_{\lambda_1\lambda_2\lambda_3}^*$-terms}. For $n=3$, the Grassmann parity of $\Theta_{\lambda_1\lambda_2\lambda_3}^*$-terms \rf{14052019-man02-15} is given in \rf{19052019-man02-12}, where $\Ebf_\lambda$ is defined in \rf{19052019-man02-08}. The $\Theta_{\lambda_1\lambda_2\lambda_3}^*$-terms having $(-)^{\Ebf_\lambda}=1$ are Grassmann even, while the $\Theta_{\lambda_1\lambda_2\lambda_3}^*$-terms having $(-)^{\Ebf_\lambda}=-1$ are Grassmann odd. Using basis of $\Phi^*$, $\Psi^*$ superfields \rf{16052019-man02-22}, we now present all Grassmann even and odd $\Theta_{\lambda_1\lambda_2\lambda_3}^*$-terms which are needed for the basis of independent vertices.
\beq
&& \hspace{-2cm} \hbox{\bf Grassmann even $\Theta^{*3}$-terms}, \qquad (-)^{\Ebf_\lambda} = 1 \quad \Longleftrightarrow \quad \hbox{$\Mbf_\lambda$ - integer};
\nonumber\\
\label{20052019-man02-01} && \hspace{-2cm} \hbox{$\Phi^{*3}$-terms}: \hspace{0.8cm} \Phi_{s_1-\half}^* \Phi_{s_2-\half}^* \Phi_{-s_3}^* \,, \hspace{0.9cm}  \Phi_{-s_1}^* \Phi_{-s_2}^* \Phi_{-s_3}^* \,,
\\
\label{20052019-man02-02} && \hspace{-2cm} \hbox{$\Phi^{*2}\Psi^*$-terms}: \hspace{0.4cm} \Phi_{s_1-\half}^* \Phi_{s_2-\half}^* \Psi_{s_3}^* \,, \hspace{1cm} \Phi_{s_1-\half}^* \Phi_{-s_2}^* \Psi_{-s_3-\half}^* \,, \hspace{0.3cm}  \Psi_{s_1}^*\Phi_{-s_2}^* \Phi_{-s_3}^*  \,,   \qquad
\\
\label{20052019-man02-03} && \hspace{-2cm} \hbox{$\Phi^*\Psi^{*2} $-terms}: \hspace{0.4cm}  \Phi_{s_1-\half}^* \Psi_{s_2}^* \Psi_{-s_3-\half}^* \,, \qquad
\Psi_{s_1}^* \Psi_{s_2}^* \Phi_{-s_3}^*\,,   \hspace{1.5cm}
\Phi_{-s_1}^* \Psi_{-s_2-\half}^* \Psi_{-s_3-\half}^* \,,
\\
\label{20052019-man02-04} && \hspace{-2cm} \hbox{$\Psi^{*3}$-terms}: \hspace{0.8cm}  \Psi_{s_1}^* \Psi_{s_2}^* \Psi_{s_3}^* \,, \hspace{2cm} \Psi_{s_1}^* \Psi_{-s_2-\half}^* \Psi_{-s_3-\half}^* \,,
\\
&& \hspace{-2cm} \hbox{\bf Grassmann odd $\Theta^{*3}$-terms}, \qquad (-)^{\Ebf_\lambda} = -1 \quad \Longleftrightarrow \quad \hbox{$\Mbf_\lambda$ - half-integer};
\nonumber\\
\label{20052019-man02-05} && \hspace{-2cm} \hbox{$\Phi^{*3}$-terms}: \hspace{0.6cm}   \Phi_{-s_1}^* \Phi_{-s_2}^* \Phi_{s_3-\half}^*,    \hspace{1.5cm} \Phi_{s_1-\half}^* \Phi_{s_2-\half}^* \Phi_{s_3-\half}^* \,,
\\
\label{20052019-man02-06} && \hspace{-2cm} \hbox{$\Phi^{*2}\Psi^*$-terms}: \hspace{0.1cm} \Phi_{-s_1}^* \Phi_{-s_2}^* \Psi_{-s_3-\half}^* \,,
\hspace{1.2cm} \Phi_{-s_1}^* \Phi_{s_2-\half}^* \Psi_{s_3}^* \,, \hspace{1.5cm}
\Psi_{-s_1-\half}^* \Phi_{s_2-\half}^* \Phi_{s_3-\half}^* \,,
\\
\label{20052019-man02-07} && \hspace{-2cm} \hbox{$\Phi^*\Psi^{*2} $-terms}: \hspace{0.1cm}  \Phi_{-s_1}^* \Psi_{-s_2-\half}^* \Psi_{s_3}^* \,,  \hspace{1.5cm}  \Psi_{-s_1-\half}^* \Psi_{-s_2-\half}^* \Phi_{s_3-\half}^* \,,  \hspace{0.3cm} \Phi_{s_1-\half}^* \Psi_{s_2}^* \Psi_{s_3}^* \,,
\\
\label{20052019-man02-08} && \hspace{-2cm} \hbox{$\Psi^{*3}$-terms}: \hspace{0.6cm}   \Psi_{-s_1-\half}^* \Psi_{-s_2-\half}^* \Psi_{-s_3-\half}^*, \hspace{0.4cm}     \Psi_{-s_1-\half}^* \Psi_{s_2}^* \Psi_{s_3}^* \,,
\eeq
where, for $\Phi^*$ superfields, $s_a\geq 1$, $a=1,2,3$,  while, for $\Psi^*$ superfields, $s_a\geq 0$, $a=1,2,3$ .

\appendix{ Derivation of cubic vertex $p_{\lambda_1\lambda_2\lambda_3}^-$  \rf{19052019-man02-01}.}

We split our procedure of the derivation of the cubic vertex $p_\smp3^-$ \rf{19052019-man02-01} into the following five steps.

\noindent {\bf Step 1}. Requirement in \rf{17052019-man02-29} implies that the cubic vertex $p_\smp3^-$ can be presented as
\be \label{18052019-man02-01}
p_\smp3^- = V(\Po^\Lsm,\Po_\theta) + \Vb(\Po^\Rsm,\Po_\theta)\,.
\ee

\noindent {\bf Step 2}. Using the expression for $q_\smp3^{-\Lsm}$ \rf{17052019-man02-27} and requiring the $q_\smp3^{-\Lsm}$ to be polynomial in $\Po^\Rsm$ \rf{17052019-man02-28}, we find that $ V(\Po^\Lsm,\Po_\theta)$ \rf{18052019-man02-01} is independent of $\Po_\theta$. Using the expression for $q_\smp3^{-\Rsm}$ \rf{17052019-man02-27}  and requiring the $q_\smp3^{-\Rsm}$ to be polynomial in $\Po^\Lsm$ \rf{17052019-man02-28}, we find that $\Vb(\Po^\Lsm,\Po_\theta)$ \rf{18052019-man02-01} is degree-1 homogeneous monomial in the Grassmann momentum $\Po_\theta$.
Thus, we have the relations
\be \label{18052019-man02-02}
V= V(\Po^\Lsm), \qquad \Vb= V(\Po^\Rsm,\Po_\theta), \qquad (\Po_\theta \partial_{\Po_\theta} -1)\Vb=0\,.
\ee

\noindent {\bf Step 3}. Using relations for $p_\smp3^-$ in \rf{18052019-man02-01},\rf{18052019-man02-02}, we find that equations \rf{17052019-man02-24},\rf{17052019-man02-25}  and \rf{17052019-man02-28} amount to the following
\beq
&& \hspace{-3cm} \hbox{ Equations for } \ V:
\nonumber\\[-5pt]
\label{18052019-man02-03} && \big(  N_{\Po^\Lsm} + \half - \half \Ebf_{\lambda+\half} + \sum_{a=1,2,3}\beta_a\partial_{\beta_a} \big) V= 0\,,
\\
\label{18052019-man02-04} && \big( - N_{\Po^\Lsm} + \Mbf_\lambda  + 1\big) V = 0\,,
\\
\label{18052019-man02-05} && \big( - \No_\beta -  \Mo_\lambda   + \half \Eo_{\lambda+\half} \big) V =  0\,.
\\
&& \hspace{-3cm}  \hbox{ Equations for } \ \Vb:
\nonumber\\[-5pt]
\label{18052019-man02-06} && \big( N_{\Po^\Rsm}+ 2 - \half \Ebf_{\lambda+\half} + \sum_{a=1,2,3}\beta_a\partial_{\beta_a} \big) \Vb = 0\,,
\\
\label{18052019-man02-07} && \big( N_{\Po^\Rsm}    + \Mbf_\lambda  + \half \big) \Vb = 0\,,
\\
\label{18052019-man02-08} && \big( - \No_\beta +  \Mo_\lambda   + \half \Eo_{\lambda+\half} \big) \Vb =  0\,,
\eeq
where $N_{\Po^\Rsm}$, $N_{\Po^\Lsm}$ and $\No_\beta$  are defined in \rf{17052019-man02-09} and \rf{17052019-man02-20} respectively and we use the notation
\be
\Mbf_\lambda \equiv \sum_{a=1,2,3} \lambda_a\,,\quad \Ebf_{\lambda+\half} \equiv \sum_{a=1,2,3} e_{\lambda_a+\half}\,, \quad  \Mo_\lambda = \frac{1}{3}\sum_{a=1,2,3}\betach_a \lambda_a\,,
\quad \Eo_{\lambda+\half} = \frac{1}{3}\sum_{a=1,2,3}\betach_a e_{\lambda_a+\half}\,.
\ee
We note that equations \rf{18052019-man02-03},\rf{18052019-man02-06} and \rf{18052019-man02-04},\rf{18052019-man02-07} are obtained from the respective equations \rf{17052019-man02-24} and \rf{17052019-man02-25}, while the equations
\rf{18052019-man02-05} and \rf{18052019-man02-08} amount to requiring the $j_\smp3^{-\Lsm}$ and $j_\smp3^{-\Rsm}$ \rf{17052019-man02-26} to be polynomial in $\Po^\Rsm,\Po^\Lsm$.
We note also that equations \rf{18052019-man02-05} and \rf{18052019-man02-08} are simply the  respective equations  $\Jbf^{-\Lsm}V=0$ and $\Jbf^{-\Rsm}\Vb=0$, where we use notation in \rf{17052019-man02-18},\rf{17052019-man02-19} .

Analysis of system of equations \rf{18052019-man02-03}-\rf{18052019-man02-05} and \rf{18052019-man02-06}-\rf{18052019-man02-08} is identical. Therefore to avoid the repetitions, we consider the system of equations  \rf{18052019-man02-03}-\rf{18052019-man02-05}.

\noindent {\bf Step 4}. We consider equation \rf{18052019-man02-04}. This equation is solved as
\be \label{18052019-man02-09}
V = (\Po^\Lsm)^{\Mbf_\lambda + 1 } V^{(1)}\,, \qquad V^{(1)} = V^{(1)}(\beta_1,\beta_2,\beta_3)\,,
\ee
where a new vertex $V^{(1)}$ depends only on the momenta $\beta_1,\beta_2,\beta_3$ and the helicities $\lambda_1,\lambda_2,\lambda_3$. Using \rf{18052019-man02-09}, we find that equations \rf{18052019-man02-03},\rf{18052019-man02-05} amount to the following two equations for the vertex $V^{(1)}$:
\beq
\label{18052019-man02-10} &&  \big( \Mbf_\lambda + \frac{3}{2} - \half \Ebf_{\lambda+\half} + \sum_{a=1,2,3}\beta_a\partial_{\beta_a} \big)V^{(1)} = 0\,,
\\
\label{18052019-man02-11} && \big( \No_\beta +  \Mo_\lambda   - \half \Eo_{\lambda+\half} \big) V^{(1)} =  0\,.
\eeq

\noindent {\bf Step 5}. We consider equations \rf{18052019-man02-10},\rf{18052019-man02-11}.
Using relation $e_{\lambda+\half} = 1 - e_\lambda$, these equations can be represented as
\be \label{18052019-man02-12}
\big(  \Mbf_{\lambda+\half e} + \sum_{a=1,2,3}\beta_a\partial_{\beta_a}\big) V^{(1)}   = 0\,, \qquad \big( \No_\beta +  \Mo_{\lambda+\half e} \big) V^{(1)} =  0\,.
\ee
Introducing a new vertex $V^{(2)}$ by the relation
\be \label{18052019-man02-13}
V^{(1)} = V^{(2)}  \prod_{a=1,2,3} \beta_a^{-\lambda_a - \half e_{\lambda_a} }\,,
\ee
we find that equations \rf{18052019-man02-12} amount to the following respective equations for the vertex $V^{(2)}$:
\be \label{18052019-man02-14}
\sum_{a=1,2,3}\beta_a\partial_{\beta_a} \,\, V^{(2)}   = 0\,, \qquad \No_\beta \,\, V^{(2)} =  0\,.
\ee
Equations \rf{18052019-man02-14} tell us that the vertex $V^{(2)}$ is independent of the momenta $\beta_1$, $\beta_2$, $\beta_3$,
\be \label{18052019-man02-15}
V^{(2)} = C^{\lambda_1\lambda_2\lambda_3} \,,
\ee
where $C^{\lambda_1\lambda_2\lambda_3}$ is a constant which depends only on the helicities.
Collecting relations in \rf{18052019-man02-09}-\rf{18052019-man02-15},  we get expression for $V_{\lambda_1\lambda_2\lambda_3}$ given in \rf{19052019-man02-02}. Repeating analysis above-given for case of $\Vb$ we find solution to $\Vb_{\lambda_1\lambda_2\lambda_3}$ given in \rf{19052019-man02-03}. Plugging expressions for $V_{\lambda_1\lambda_2\lambda_3}$, $\Vb_{\lambda_1\lambda_2\lambda_3}$ \rf{19052019-man02-02}, \rf{19052019-man02-03} into \rf{17052019-man02-26},\rf{17052019-man02-27}, we find expressions for $q_{\lambda_1\lambda_2\lambda_3}^{-\Rsm,\Lsm}$ and $j_{\lambda_1\lambda_2\lambda_3}^{-\Rsm,\Lsm}$ given in \rf{19052019-man02-04}-\rf{19052019-man02-07}. Note that while deriving expressions for $q_\smp3^{-\Rsm,\Lsm}$ \rf{19052019-man02-04},\rf{19052019-man02-05} from relations in \rf{17052019-man02-27}, we used restrictions on $\epsilon$ \rf{17052019-man02-17-a1} given in \rf{19052019-man02-09},\rf{19052019-man02-10}.



\small
\begin{thebibliography}{30}

\parskip=-5.pt







\bibitem{Dirac:1949cp}
  P.~A.~M.~Dirac,
  Rev.\ Mod.\ Phys.\  {\bf 21}, 392 (1949).



\bibitem{Brink:1982wv}
  L.~Brink, O.~Lindgren and B.~E.~W.~Nilsson,
  Phys.\ Lett.\  {\bf 123B}, 323 (1983).

\bibitem{Mandelstam:1982cb}
  S.~Mandelstam,
  Nucl.\ Phys.\ B {\bf 213}, 149 (1983).




 \bibitem{Green:1983hw}
  M.~B.~Green, J.~H.~Schwarz and L.~Brink,
  Nucl.\ Phys.\ B {\bf 219}, 437 (1983).
%
\\
%
  M.~B.~Green and J.~H.~Schwarz,
  Nucl.\ Phys.\ B {\bf 218}, 43 (1983).
%
\\
%
  M.~B.~Green and J.~H.~Schwarz,
  Nucl.\ Phys.\ B {\bf 243}, 475 (1984).


\bibitem{Bergman:1995wh}
  O.~Bergman and C.~B.~Thorn,
  Phys.\ Rev.\ D {\bf 52}, 5980 (1995)
  [hep-th/9506125].


\bibitem{deWit:1988wri}
  B.~de Wit, J.~Hoppe and H.~Nicolai,
  Nucl.\ Phys.\ B {\bf 305}, 545 (1988).
%
\\
%
  E.~Bergshoeff, E.~Sezgin, Y.~Tanii and P.~K.~Townsend,
  Annals Phys.\  {\bf 199}, 340 (1990).




\bibitem{Metsaev:2017cuz}
  R.~R.~Metsaev,
  JHEP {\bf 1711}, 197 (2017)
  [arXiv:1709.08596 [hep-th]].

\bibitem{Metsaev:2018moa}
  R.~R.~Metsaev,
  JHEP {\bf 1812}, 055 (2018)
  [arXiv:1809.09075 [hep-th]].

\bibitem{Brodsky:2013dca}
  S.J.Brodsky, G.F.de Téramond and H.G.Dosch,
  Few Body Syst.\  {\bf 55}, 407 (2014)
  [arXiv:1310.8648].
%
\\
%
  S.~J.~Brodsky, G.~F.~de Teramond, H.G.Dosch, J.Erlich,
  Phys.\ Rept.\  {\bf 584}, 1 (2015)
  [arXiv:1407.8131]


\bibitem{Siegel:1988yz}
  W.~Siegel,
  ``Introduction To String Field Theory,''
  arXiv:hep-th/0107094.




\bibitem{Metsaev:2017myp}
  R.~R.~Metsaev,
  J.\ Phys.\ A {\bf 51}, no. 21, 215401 (2018)
  [arXiv:1711.11007 [hep-th]].


\bibitem{Metsaev:2019opn}
  R.~R.~Metsaev,
  Phys.\ Lett.\ B {\bf 793}, 134 (2019)
  [arXiv:1903.10495 [hep-th]].



\bibitem{Bengtsson:1983pd}
  A.~K.~H.~Bengtsson, I.~Bengtsson and L.~Brink,
  Nucl.\ Phys.\ B {\bf 227}, 31 (1983).


\bibitem{Bengtsson:1983pg}
  A.~K.~H.~Bengtsson, I.~Bengtsson and L.~Brink,
  Nucl.\ Phys.\ B {\bf 227}, 41 (1983).


\bibitem{Bengtsson:1986kh}
  A.~K.~H.~Bengtsson, I.~Bengtsson and N.~Linden,
  Class.\ Quant.\ Grav.\  {\bf 4}, 1333 (1987).






\bibitem{Metsaev:2005ar}
  R.~R.~Metsaev,
  Nucl.\ Phys.\ B {\bf 759}, 147 (2006)
  [hep-th/0512342].



\bibitem{Metsaev:2007rn}
  R.~R.~Metsaev,
  Nucl.\ Phys.\ B {\bf 859}, 13 (2012)
  [arXiv:0712.3526 [hep-th]].


\bibitem{Green:1982tk}
  M.~B.~Green and J.~H.~Schwarz,
  Phys.\ Lett.\  {\bf 122B}, 143 (1983).

\bibitem{Ananth:2015tsa}
  S.~Ananth, L.~Brink and M.~Mali,
  JHEP {\bf 1508}, 153 (2015)
  [arXiv:1507.01068 [hep-th]].


\bibitem{Metsaev:2004wv}
  R.~R.~Metsaev,
  Phys.\ Rev.\ D {\bf 71}, 085017 (2005)
  [hep-th/0410239].



\bibitem{Buchbinder:2017nuc}
  I.~L.~Buchbinder, S.~J.~Gates and K.~Koutrolikos,
  Universe {\bf 4}, no. 1, 6 (2018)
  [arXiv:1708.06262].


\bibitem{Koutrolikos:2017qkx}
  K.~Koutrolikos, P.~Kočí and R.~von Unge,
  JHEP {\bf 1803}, 119 (2018)
  [arXiv:1712.05150 [hep-th]].



\bibitem{Buchbinder:2018wwg}
  I.~L.~Buchbinder, S.~J.~Gates and K.~Koutrolikos,
  JHEP {\bf 1805}, 204 (2018)
  [arXiv:1804.08539 [hep-th]].


\bibitem{Buchbinder:2018wzq}
  I.~L.~Buchbinder, S.~J.~Gates and K.~Koutrolikos,
  JHEP {\bf 1808}, 055 (2018)
  [arXiv:1805.04413 [hep-th]].

\bibitem{Buchbinder:2018nkp}
  E.~I.~Buchbinder, J.~Hutomo and S.~M.~Kuzenko,
  JHEP {\bf 1809}, 027 (2018)
  [arXiv:1805.08055 [hep-th]].



\bibitem{Gates:2019cnl}
  S.~J.~Gates and K.~Koutrolikos,
  ``Progress on cubic interactions of arbitrary superspin supermultiplets via gauge invariant supercurrents,''
  arXiv:1904.13336 [hep-th].


\bibitem{Kuzenko:1993jq}
  S.~M.~Kuzenko and A.~G.~Sibiryakov,
  JETP Lett.\  {\bf 57}, 539 (1993)
%
\\
%
  S.~M.~Kuzenko, A.~G.~Sibiryakov and V.~V.~Postnikov,
  JETP Lett.\  {\bf 57}, 534 (1993)



\bibitem{Conde:2016izb}
  E.~Conde, E.~Joung and K.~Mkrtchyan,
  JHEP {\bf 1608}, 040 (2016)
  [arXiv:1605.07402 [hep-th]].




\bibitem{Fotopoulos:2010ay}
  A.~Fotopoulos and M.~Tsulaia,
  JHEP {\bf 1011}, 086 (2010)
  [arXiv:1009.0727 [hep-th]].
%
\\
%
  R.~Manvelyan, K.~Mkrtchyan and W.~Ruhl,
  Nucl.\ Phys.\ B {\bf 836}, 204 (2010)
  [arXiv:1003.2877 [hep-th]].
%
\\
  A.~Sagnotti and M.~Taronna,
  Nucl.\ Phys.\  B {\bf 842}, 299 (2011)
  [arXiv:1006.5242 [hep-th]].
%
\\
%
  R.~Manvelyan, K.~Mkrtchyan and W.~Ruehl,
  Phys.\ Lett.\  B {\bf 696}, 410 (2011)
  [arXiv:1009.1054 [hep-th]].


\bibitem{Metsaev:2012uy}
  R.~R.~Metsaev,
  Phys.\ Lett.\ B {\bf 720}, 237 (2013)
  [arXiv:1205.3131 [hep-th]].


\bibitem{Konstein:1989ij}
  S.~E.~Konstein and M.~A.~Vasiliev,
  Nucl.\ Phys.\ B {\bf 331}, 475 (1990).
%
\\
%
  S.~E.~Konshtein and M.~A.~Vasiliev,
  Nucl.\ Phys.\ B {\bf 312}, 402 (1989).



\bibitem{Engquist:2002vr}
  J.~Engquist, E.~Sezgin and P.~Sundell,
  Class.\ Quant.\ Grav.\  {\bf 19}, 6175 (2002)
  [hep-th/0207101].


\bibitem{Metsaev:1991mt}
  R.~R.~Metsaev,
  Mod.\ Phys.\ Lett.\ A {\bf 6}, 359 (1991).

\bibitem{Metsaev:1991nb}
  R.~R.~Metsaev,
  Mod.\ Phys.\ Lett.\ A {\bf 6}, 2411 (1991).



\bibitem{Ponomarev:2016lrm}
  D.~Ponomarev and E.~D.~Skvortsov,
  J.\ Phys.\ A {\bf 50}, no. 9, 095401 (2017)
  [arXiv:1609.04655 [hep-th]].




\bibitem{Henneaux:2012wg}
  M.~Henneaux, G.~L.~Gomez and R.~Rahman,
  JHEP {\bf 1208}, 093 (2012)
  [arXiv:1206.1048 [hep-th]].
%
\\
  M.~Henneaux, G.~L.~Gomez and R.~Rahman,
  JHEP {\bf 1401}, 087 (2014)  [arXiv:1310.5152 [hep-th]].


\bibitem{Akshay:2015kxa}
  Y.~S.~Akshay and S.~Ananth,
  Phys.\ Rev.\ D {\bf 91}, 085029 (2015)
  [arXiv:1504.00967 [hep-th]].



\bibitem{Najafizadeh:2018cpu}
  M.~Najafizadeh,
  Phys.\ Rev.\ D {\bf 98}, no. 12, 125012 (2018)
  [arXiv:1807.01124 [hep-th]].

\bibitem{Bekaert:2005jf}
  X.~Bekaert, N.~Boulanger and S.~Cnockaert,
  JHEP {\bf 0601}, 052 (2006)
  [hep-th/0508048].


\bibitem{Kaparulin:2019quz}
  D.~S.~Kaparulin and S.~L.~Lyakhovich,
   ``A note on unfree gauge symmetry,''
  arXiv:1904.04038 [hep-th].




\bibitem{Sorokin:2018djm}
  D.~Sorokin and M.~Tsulaia,
  Nucl.\ Phys.\ B {\bf 929}, 216 (2018)
  [arXiv:1801.04615 [hep-th]].
%
\\
%
  V.~K.~Dobrev,
  Nucl.\ Phys.\ B {\bf 854}, 878 (2012)
  [arXiv:1012.3685 [hep-th]].






\bibitem{Buchbinder:2015kca}
  I.~L.~Buchbinder and K.~Koutrolikos,
  JHEP {\bf 1512}, 106 (2015)
  [arXiv:1510.06569 [hep-th]].




\bibitem{Zinoviev:2007js}
  Y.~M.~Zinoviev,
  Nucl.\ Phys.\ B {\bf 785}, 98 (2007)
  [arXiv:0704.1535 [hep-th]].



\bibitem{Vasiliev:1990en}
  M.~A.~Vasiliev,
  Phys.\ Lett.\ B {\bf 243}, 378 (1990).



\bibitem{Alkalaev:2002rq}
  K.~B.~Alkalaev and M.~A.~Vasiliev,
  Nucl.\ Phys.\ B {\bf 655}, 57 (2003)
  [hep-th/0206068].
%
\\
%
  K.~Alkalaev,
  JHEP {\bf 1103}, 031 (2011)
  [arXiv:1011.6109 [hep-th]].

\bibitem{Metsaev:2018xip}
  R.~R.~Metsaev,
  Nucl.\ Phys.\ B {\bf 936}, 320 (2018)
  [arXiv:1807.07542 [hep-th]].

\bibitem{Joung:2011ww}
  E.~Joung and M.~Taronna,
  Nucl.\ Phys.\ B {\bf 861}, 145 (2012)
  [arXiv:1110.5918 [hep-th]].
%
\\
%
  E.~Joung, L.~Lopez and M.~Taronna,
  JHEP {\bf 1301}, 168 (2013)
  [arXiv:1211.5912 [hep-th]].
%
\\
%
  E.~Joung, L.~Lopez and M.~Taronna,
  J.\ Phys.\ A {\bf 46}, 214020 (2013)
  [arXiv:1207.5520 [hep-th]].
%
\\
%
  C.~Sleight,
  J.\ Phys.\ A {\bf 50}, no. 38, 383001 (2017)
  [arXiv:1610.01318 [hep-th]].


\bibitem{Alkalaev:1999hi}
  K.~B.~Alkalaev and S.~L.~Lyakhovich,
  Mod.\ Phys.\ Lett.\ A {\bf 14}, 2727 (1999).
%
\\
%
  D.~V.~Uvarov,
  J.\ Phys.\ A {\bf 51}, no. 28, 285402 (2018)
  [arXiv:1707.05761 [hep-th]].
%
\\
%
  D.~V.~Uvarov,
  Eur.\ Phys.\ J.\ C {\bf 79}, no. 5, 425 (2019).




\bibitem{Nilsson:2015pua}
  B.~E.~W.~Nilsson,
  JHEP {\bf 1608}, 142 (2016)
  [arXiv:1506.03328 [hep-th]].



\bibitem{Basile:2017mqc}
  T.~Basile, R.~Bonezzi and N.~Boulanger,
  JHEP {\bf 1704}, 054 (2017)
  [arXiv:1701.08645 [hep-th]].



\bibitem{Buchbinder:2017izy}
  I.~L.~Buchbinder, T.~V.~Snegirev and Y.~M.~Zinoviev,
  JHEP {\bf 1708}, 021 (2017)
  [arXiv:1705.06163].



\bibitem{Henneaux:2018agj}
  M.Henneaux, V.Lekeu, A.Leonard, J.Matulich, S.Prohazka,
  JHEP {\bf 1811}, 156 (2018)
  arXiv:1810.04457



\bibitem{Kuzenko:2018lru}
  S.~M.~Kuzenko and M.~Ponds,
  JHEP {\bf 1810}, 160 (2018)
  [arXiv:1806.06643 [hep-th]].

\bibitem{Kuzenko:2016qwo}
  S.~M.~Kuzenko and M.~Tsulaia,
  Nucl.\ Phys.\ B {\bf 914}, 160 (2017)
  [arXiv:1609.06910 [hep-th]].

\bibitem{Kessel:2018ugi}
  P.~Kessel and K.~Mkrtchyan,
  Phys.\ Rev.\ D {\bf 97}, no. 10, 106021 (2018)
  [arXiv:1803.02737 [hep-th]].
%
\\
%
  S.~Fredenhagen, O.~Kruger and K.~Mkrtchyan,
 ``Vertex-Constraints in 3D Higher Spin Theories,''
  arXiv:1905.00093 [hep-th].



\bibitem{Metsaev:1999ui}
  R.~R.~Metsaev,
  Nucl.\ Phys.\ B {\bf 563}, 295 (1999)
  [arXiv:hep-th/9906217].

\bibitem{Metsaev:2000qb}
  R.~R.~Metsaev,
  Nucl.\ Phys.\ Proc.\ Suppl.\  {\bf 102}, 100 (2001)
  [hep-th/0103088].


\bibitem{Metsaev:2016rpa}
R.~R.~Metsaev,
``Interacting light-cone gauge conformal fields,''
  arXiv:1612.06348 [hep-th].




\bibitem{Skvortsov:2018jea}
  E.~D.~Skvortsov, T.~Tran, M.~Tsulaia,
  Phys.\ Rev.\ Lett.\  {\bf 121}, no. 3, 031601 (2018)
  [arXiv:1805.00048].



\bibitem{Ponomarev:2016jqk}
  D.~Ponomarev and A.~A.~Tseytlin,
  JHEP {\bf 1605}, 184 (2016)
  [arXiv:1603.06273 [hep-th]].











\end{thebibliography}
\end{document}